\documentclass[aps,pra,tightenlines,superscriptaddress,floatfix]{revtex4}
\usepackage[utf8]{inputenc}
\usepackage[T1]{fontenc}

\usepackage{amssymb}
\usepackage{graphicx}
\usepackage{amsmath}
\usepackage{amsbsy}
\usepackage{amsthm}
\usepackage{bbm}
\usepackage{epsfig}
\usepackage{color}

\newcommand{\beq}{\begin{eqnarray}}
\newcommand{\eeq}{\end{eqnarray}}

\begin{document}

\title{Free versus Bound Entanglement: Machine learning tackling a NP-hard problem}

\author{Beatrix C. Hiesmayr}
\email{Beatrix.Hiesmayr@univie.ac.at}

\affiliation{University of Vienna, Faculty of Physics, Währingerstrasse 17, 1090 Vienna, Austria.}

\begin{abstract}
Entanglement detection in high dimensional systems is a NP-hard problem since it is lacking an efficient way.  Given a bipartite quantum state of interest \textit{free} entanglement can be detected efficiently by the PPT-criterion (Peres-Horodecki criterion), in contrast to detecting \textit{bound} entanglement, i.e. a curious form of entanglement that can also not be distilled into maximally (free) entangled states. Only a few bound entangled states have been found, typically by constructing dedicated entanglement witnesses, so naturally the question arises how large is the volume of those states. We define a large family of magically symmetric states of bipartite qutrits for which we find $82\%$ to be free entangled, $2\%$ to be certainly separable and as much as $10\%$ to be bound entangled, which shows that this kind of entanglement is not rare. Via various machine learning algorithms we can confirm that the remaining $6\%$ of states are more likely to belonging to the set of separable states than bound entangled states. Most important we find via dimension reduction algorithms that there is a strong $2$-dimensional (linear) sub-structure in the set of bound entangled states. This revealed structure opens a novel path to find and characterize bound entanglement towards solving the long-standing problem of what the existence of bound entanglement is implying.
\end{abstract}

\flushbottom
\maketitle
\thispagestyle{empty}

\section{Introduction}

\begin{figure}[h!]
\begin{center}
\includegraphics[width=13cm,keepaspectratio=true]{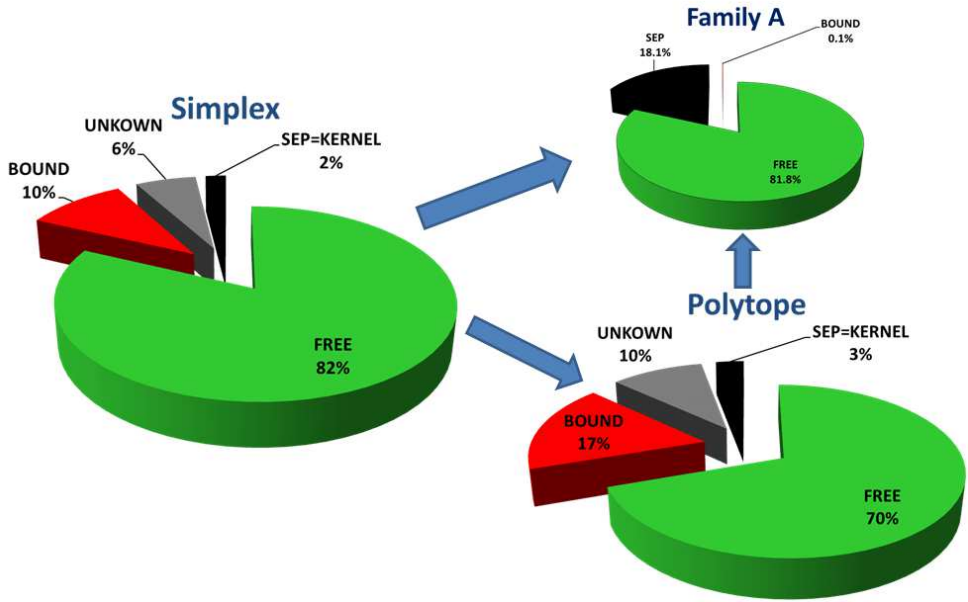}
\caption{This shows the final result of occurrences of states based on the chosen grading of $\Delta=\frac{1}{18}$ for the simplex $\mathcal{W}$ and polytope $\mathcal{P}$ and $\Delta=\frac{1}{72}$ for the family $A$.}\label{simpelexsummary}
\end{center}
\end{figure}

Entanglement is known to be the resource to outperform algorithms running on classical physical systems. Entanglement has been shown to exists in many physical systems at low and high energies and for instance is also currently explored for its potential to detect cancer in human beings~\cite{PETcancer1,PETcancer2,PETcancer3}. Even given the full available information, i.e. the quantum state of a bipartite system, no general efficient method is known to detect entanglement except for low-dimensional system such as $2\otimes2$ and $2\otimes3$. Only in these dimensions can entanglement be faithfully detected by the Peres-Horodecki criterion~\cite{PPT1,PPT2}, also known as the Positive Partial Transpose (PPT) criterion, which states that if the partial transpose of a given bipartite state has at least one negative eigenvalue, the state is entangled. More precisely, entanglement detected by the PPT-criterion is often called \textit{free} entanglement since this entanglement can be distilled~\cite{firstbound}, i.e. via local operations and classical communications (LOCC) two parties can generate maximally entangled states, so called Bell states. For those states that are entangled, but the PPT criterion fails, Bell states cannot be distilled that are therefore called \textit{bound} entangled states or PPT-entangled states. Let us remark, it has not yet been proven, if there exists non-distillable states that have some strictly negative eigenvalues after partial transpose~\cite{FiveOpen}. So curiously, states exists that can be produced by mixing maximally entangled Bell states, but those states kind of absorb the entanglement in an irreversible way, i.e. the process cannot be reversed. Up to now it is not yet fully understood why those kind of states, being not free, exist and what role they play with respect to quantum information theoretic tasks.

The first example of a bound entangled state was found by the Horodecki family in the year 1998~\cite{firstbound}. Most works~\cite{Darek,Darek2,slater1,slater2,slater3,slater4,Sanperabound,Bennett,Krammer,Lokard,Bruss,HiesmayrLoeffler2,activation} construct carefully non-decomposable entanglement witnesses (decomposible witnesses do not break the PPT criterion) or use other sophisticated constructions and apply them to exemplary states employing additional symmetries. Some works have focussed on the usefulness in quantum information tasks such as whether bound entangled states can violate a Bell inequality~\cite{Vertesi} or its usefulness in quantum key generation~\cite{keybound}. In 2013 the first successful experimental observation of bipartite bound entanglement was realizes with two twisted photons, i.e. two physical photons entangled in three degrees of their orbital angular momentum~\cite{Hiesmayrboundexperiment}.

In this paper we exploit magically symmetric states, i.e. states that are convex combinations of a complete set of Bell states and its Hilbert space is equivalent to a $d^2-1$-dimensional simplex in the real space. Within this state space in dimension $d=3$ we are able to classify iteratively the states according to free, bound, separable states or states for which all our exploited methods fail, denoted as UNKNOWN. Via the help of machine learning algorithms we can show that the remaining UNKNOWN states are more consistent in being separable states than bound entangled states. Anyhow, the volume occupied by bound entangled states is larger than the one for separable states which was not to be expected from the known examples within the magic simplex~\cite{simplex1,simplex2,simplex3} analysed via dedicated entanglement witnesses.

Entanglement witnesses are observables that can detect entanglement, including bound entanglement. They are usually constructed by human intuition exploiting symmetries allowing the proof that they are bounded by all separable states. Such examples are e.g. witnesses based on mutually unbiased bases (MUBs)~\cite{MUB1}, which  exploits Bohr's complementarity, or the quasi-pure approximation~\cite{quasipure} of a state, i.e. looking for the closed pure state for which the entanglement can be derived. Those witnesses detect successfully bound entanglement also within the magic simplex, however, for one family of states within the simplex one witness is successful, the other fails and vice versa, which shows one of the main difficulties in detecting entanglement: witnesses have only limited range. Moreover, the witnesses have to be carefully adapted to the symmetry of the considered family, which causes huge problems in praxis. Consequently, given a family of states it needs a lot of human manpower and human intuition to adapt or generate a successful entanglement witness for a given family of states. To overcome this limiting factor one can generate via the known symmetries sets of equivalent families or/and witnesses but by that one generates a lot of data sets.

On the other hand machine learning (ML) algorithms are known to be powerful tools to analyse huge data sets. Therefore, to generate the ground truth we lay a grid over the magic simplex and store for all vectors their labels: FREE, BOUND, SEP or UNKNOWN if all methods fail. In addition machine learning algorithms are in general powerful tools to reveal unknown structures. Indeed, the ML algorithms find in our case a strong structure within the set of detected bound entangled states. This paves a novel way to tackle the detection of bound entangled states also for $d>3$.

The paper is organized as follows: We start by defining the magic simplex and its most important symmetries. For two families, corresponding to different slices via the $8$-dimensional magic simplex, we show the power of different analytical entanglement witnesses to detect bound entanglement and show the exemplary region of states that are certainly separable and those that are certainly free entangled. Between those two regions the states can be either free, bound or separable. Then we lay a grid over the simplex and in applying different analytical methods to detect entanglement/separability we can label the vectors according to FREE, BOUND, SEP or UNKNOWN. We show that the analytical methods leave too many vectors to be UNKNOWN rendering a further discussion meaningless. Therefore, numerical entanglement witnesses are generated exploiting the group structure within the magic simplex, which leaves finally only $6\%$ of the simplex to be UNKNOWN. Exploiting machine learning methods we show that those states are more likely to be separable states. This ground truth is employed for supervised machine learning methods and applied to our test families. In a last step we applied dimension reduction algorithms, unsupervised ML algorithms, and found a curious quantization in the first and second principal component for the data labelled BOUND, in contrast to the other ones, which leaves us with a recipe to construct extremal bound entangled states.

\section{The Magic Simplex and Entanglement Detection}

Here we introduce the state space under consideration and methods to detect entanglement and separability and introduce important test families.

\subsection{Definition: Magic Simplex}

In mathematics a $k$-simplex is a k-dimensional polytope which is composed of the convex hull of its $k + 1$ vertices. More formally, given the $k + 1$ points $p_0,\dots,p_k\in \mathbb{R}^k$ are affinely independent, which means $p_1-p_0,p_2-p_0,\dots,p_k-p_0$ are linearly independent, then the simplex is defined by
\beq
\mathcal{S}^{(k)}:=\left\{ \sum_{i=0}^{k}c_i\; p_i\;\vert\; \sum c_i=1\quad\textrm{and}\quad c_i\geq 0\right\}\;.
\eeq
Thus a $0$-simplex is a point, a $1$-simplex a line, a $2$-simplex is a triangle, a $3$-simplex is a tetrahedron and so on.
This idea has been transferred to quantum mechanics by the authors of Refs.~\cite{simplex1,simplex2,simplex3} for bipartite qudit states and for multipartite qudit states~\cite{simplex4,simplex5}. A simplex for bipartite qudit systems is defined by
\beq
\mathcal{W}^{(d)}&:=&\left\lbrace\rho=\sum_{k,l=0}^{d-1} c_{k,l}\; P_{k,l}\;\mid\;\quad  \sum_{k,l=0}^{d-1} c_{k,l}=1\quad \textrm{and}\quad c_{k,l}\geq 0\right\rbrace\;\;
\eeq
and referred to as the \textbf{magic simplex} since in dimension $d=2$ it utilizes the magic state basis introduced by Wootters and Hill~\cite{WoottersHill} allowing the analytical computation of entanglement of formation. Here the $P_{k,l}$ form an orthonormal basis of Bell states, generated by the arbitrary choice of one Bell state, e.g. $P_{0,0}=\frac{1}{d}\sum_{i,j=0}^{d-1} |ii\rangle\langle jj|$, and applying in one subsystem the Weyl operators $W_{k,l}=\sum_{j=0}^{d-1} \omega^{j\cdot k}\;|j\rangle\langle (j+l)\mod_d|$
with $\omega=e^{\frac{2\pi i}{d}}$ ($d$ root of unity), namely $P_{k,l}=W_{k,l}\otimes \mathbbm{1}_{d}\; P_{0,0}\;W_{k,l}^\dagger\otimes \mathbbm{1}_{d}$. All states are locally maximally mixed, namely the partial trace results in the total mixed state, differently stated all correlations are in the joint system. The only pure states in this family of states are the $d^2$ maximally entangled Bell states $P_{k,l}$. Except for $d=2$ it does not include all locally maximally mixed states~\cite{simplex1}. Let us also remark that the magic simplex is well defined for an orthonormal set of Bell states, however, different sets do not always form unitary equivalent magic simplexes~\cite{simplex2} for $d>2$.

\begin{figure}[t]
\begin{center}
\includegraphics[width=15cm,keepaspectratio=true]{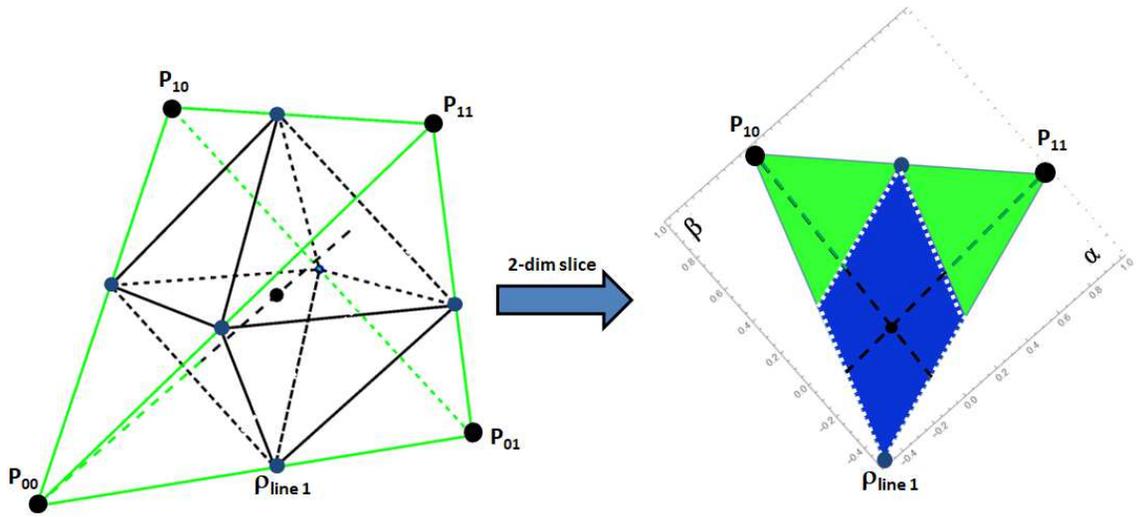}
\caption{Simplex of two qubits ($d=2$): The positivity relation requires all points representing states to be within the (green) tetrahedron. The vertices correspond to the four Bell states $P_{k,l}$, the origin to the totally maximally mixed state and all $6$ lines states $\rho_{\textrm{lines }\alpha}$ (blue dots) form a double pyramid.  The $PPT$ criterion detects all entangled states to be outside the double pyramid, which represents the kernel polytope $\mathcal{K}$ as well as the enclosure polytope $\mathcal{P}$ since there exists no bound entanglement in $d=2$. Note that $4$ out of the $8$ surface planes of the double pyramid, correspond to the optimal entanglement witnesses, while the remaining ones overlap with the tetrahedron (positivity condition). The right hand side shows a $2$-dimensional slice of the $3$-dimensional simplex. The green-black dotted line corresponds to one of the four famous isotropic lines, $\rho_{iso}=(1-p) \frac{1}{4}\mathbbm{1}+ p P_{k,l}$, for which the back part represents separable states (within $\mathcal{P}\equiv\mathcal{K}$) and the green part entangled states (PPT criterion fails). Fig.~\ref{families}~(a) shows a similar slice via the $8$-dimensional simplex of two qutrits ($d=3$).}\label{simplexqubits}
\end{center}
\end{figure}

The Hilbert space of those magically symmetric states can be visualized as a simplex in the real space $\mathbb{R}^{d^2-1}$, in particular for $d=2$ it forms the famous magic tetrahedron visualized in Fig.~\ref{simplexqubits}. All points within the tetrahedron correspond to a locally maximally mixed state and the vertices correspond to the four Bell states. Via the PPT criterion one finds all points within the double-pyramid to be separable and thus all states outside to be entangled. This solves the separability problem fully for this case.

\subsection{Structures in the magic simplex: the \textit{kernel} and \textit{enclosure} polytope}

Obviously, concerning the separability or entanglement property any of the $d^2$ Bell states are equivalent and, subsequently, certain mixtures of Bell states are with respect to entanglement equivalent. For example in dimension $d=2$ the equal mixture of any two Bell states results in an separable state, thus the middle of the line connecting two vertices ($\equiv$ Bell states) has to represent a separable state (in Fig.~\ref{simplexqubits} those six separable states are marked by blue dots and correspond to $\rho_{\textrm{lines}\alpha}$ defined below). In general, for $d=2$ the  $PPT$ criterion is necessary but also sufficient to detect entanglement, there exists only \textit{free} entanglement. For all points within the simplex it corresponds to a sign change of $y$-coordinate, i.e. it forms a double pyramid within the tetrahedron as depicted in Fig.~\ref{simplexqubits}, where the vertices are the equal mixtures of two Bell states ($\rho_{\textrm{lines }\alpha}$).

For dimensions $d>2$ the set of separable states is no longer a simple polytope within the magic simplex, however, in Ref.~\cite{simplex2} it has been proven that two polytopes can be constructed, the so called \textbf{kernel polytope $\mathcal{K}$} and the \textbf{enclosure polytope $\mathcal{P}$}, for which the authors proved that states within are separable and outside \textit{free} entangled, respectively:
\beq
\textrm{Kernel polytope:}\qquad
\mathcal{K}&:=&\left\{\rho=\sum_{\textrm{lines }\alpha} \lambda_\alpha\; \rho_{\textrm{lines }\alpha}\;\mid\;\lambda_\alpha \geq 0, \sum\lambda_\alpha=1\right\}\;.\\
\textrm{Enclosure polytope:}\qquad
\mathcal{P}&:=&\left\{\rho=\sum_{k,l=0}^{d-1} c_{k,l}\; P_{k,l}\;\mid\;c_{k,l}\in[0,\frac{1}{d}]\right\}\;.
\eeq
Thus in the region between $\mathcal{K}$ and $\mathcal{P}$ states can be either separable, bound or free entangled. The free entangled region can be detected by the PPT criterion, which does not in general form a polytope. The remaining states are either bound or separable. The kernel polytope $\mathcal{K}$ relies on the structure of so called ``\textit{lines}'' in the momentum-position space. The Weyl operators generating the maximal entangled Bell states $P_{k,l}$ imply a group structure within the magic simplex $\mathcal{W}$ such that the classical phase space $\{k,l\}$ forms a two dimensional torus with a ``\textit{linear}'' structure -- multiplication by constants and additions always done with the ring $\mathbb{Z}_d$. Therefore, each cyclic subgroup is a \textit{line} in the phase space, i.e. starting with a point $(p,q)$ a line is defined by the set $\{p+n k,q+nl\}$ with $n$ being an integer; for instance, \beq
\rho_{\textrm{lines }}=\frac{1}{d}\sum_{i=0}^{d-1} P_{0,i}\;.\label{linestates}\eeq Note that in higher dimensions there are also sublattices. The number of lines (or sublattices) with $d$ points is given by $N(d)=d(d+1+\sum b)$ where the sum runs over all $b$ that are proper divisors. For the case $d=3$ there are $12$ lines such that their convex combination forms the kernel polytope $\mathcal{K}$ and the $12$ vertices belonging to the $\binom{9}{3}=84$ $2$-faces. Thus all convex combinations of three out of nine Bell states form the polytope $\mathcal{P}$ with $84$ vertices as is discussed in detail in Ref.~\cite{simplex2}.

\begin{figure*}[t]
\begin{center}
(a)\includegraphics[width=3.7cm,keepaspectratio=true]{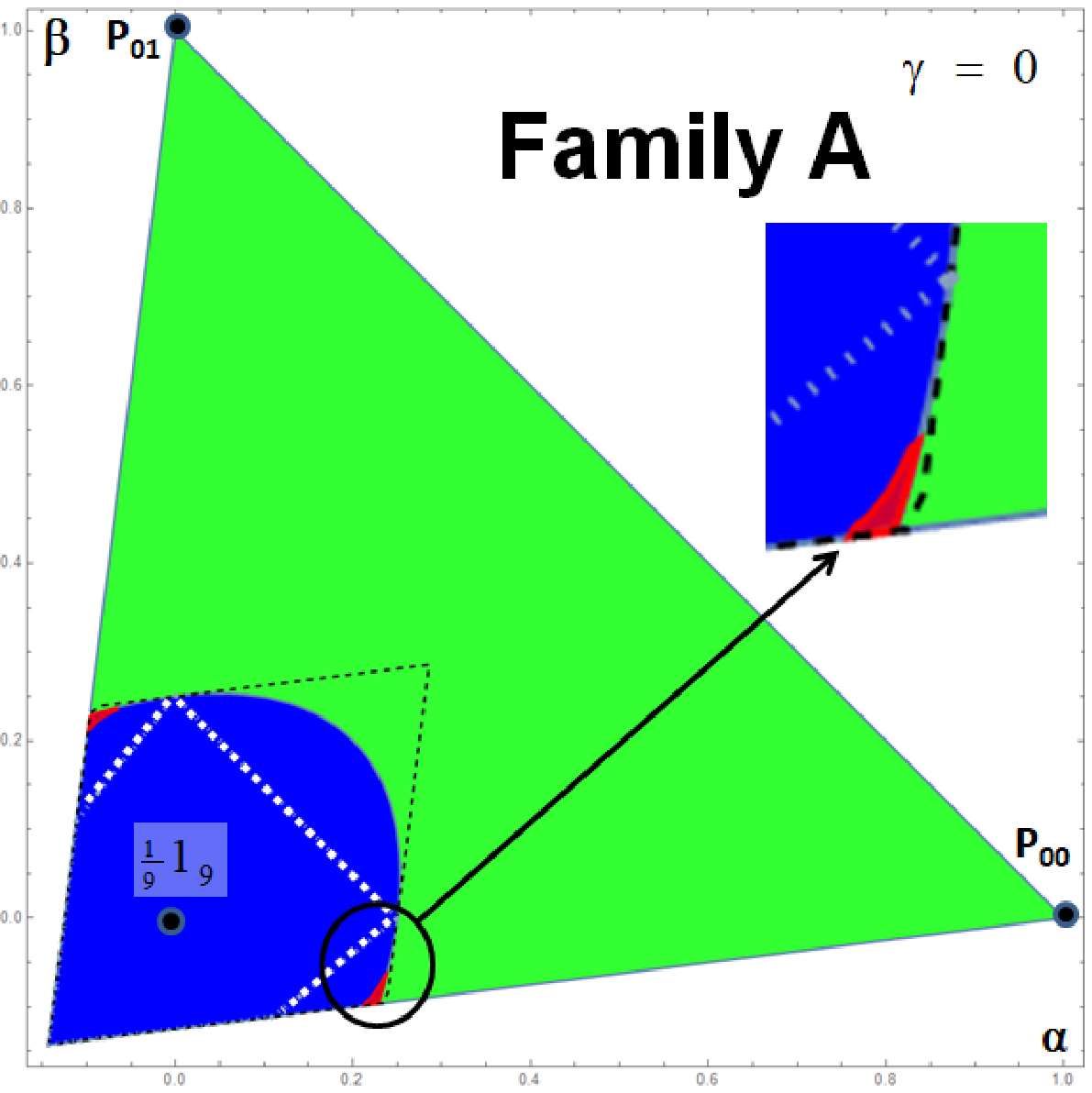}
(b)\includegraphics[width=3.7cm,keepaspectratio=true]{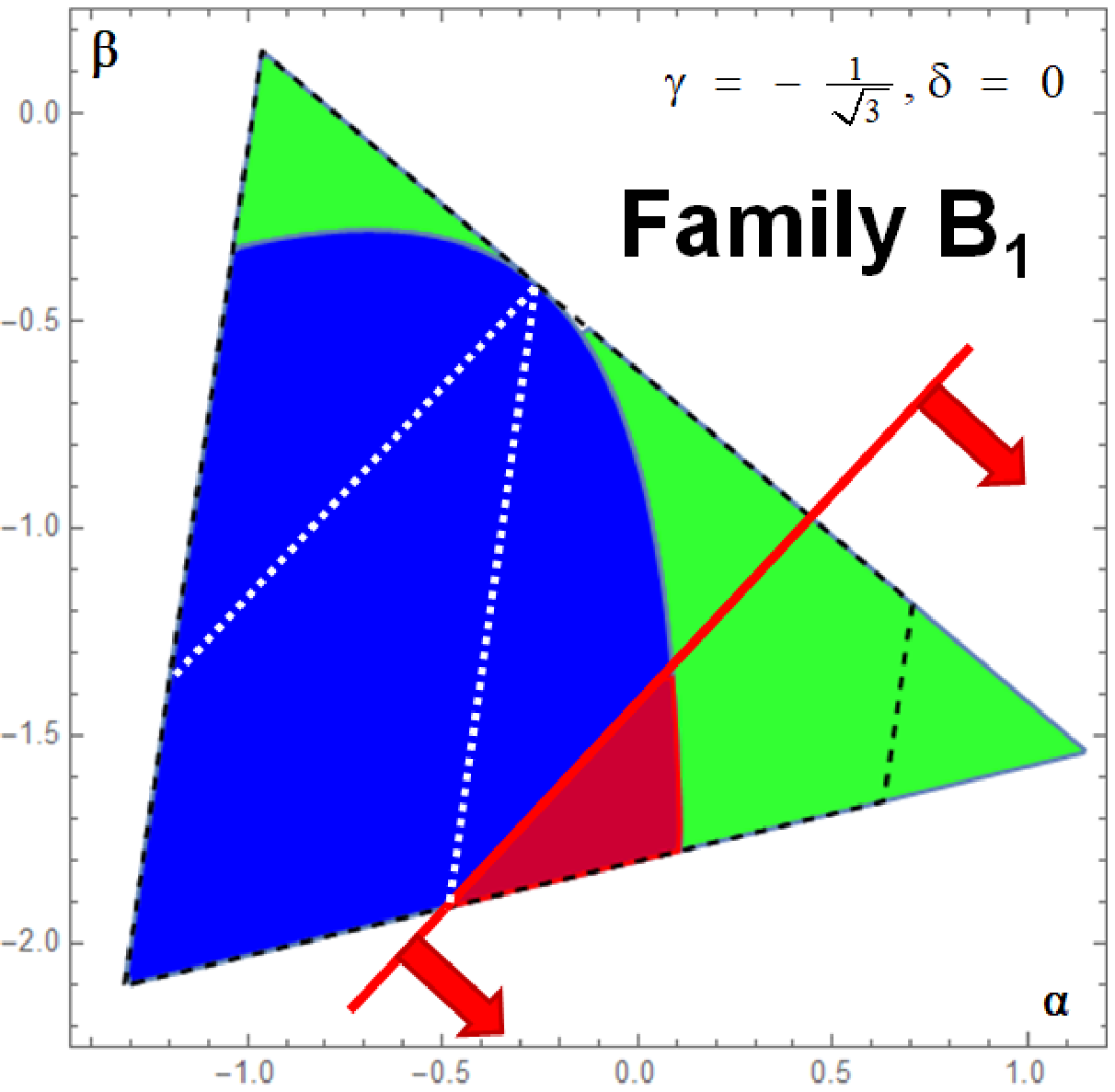}
(c)\includegraphics[width=3.7cm,keepaspectratio=true]{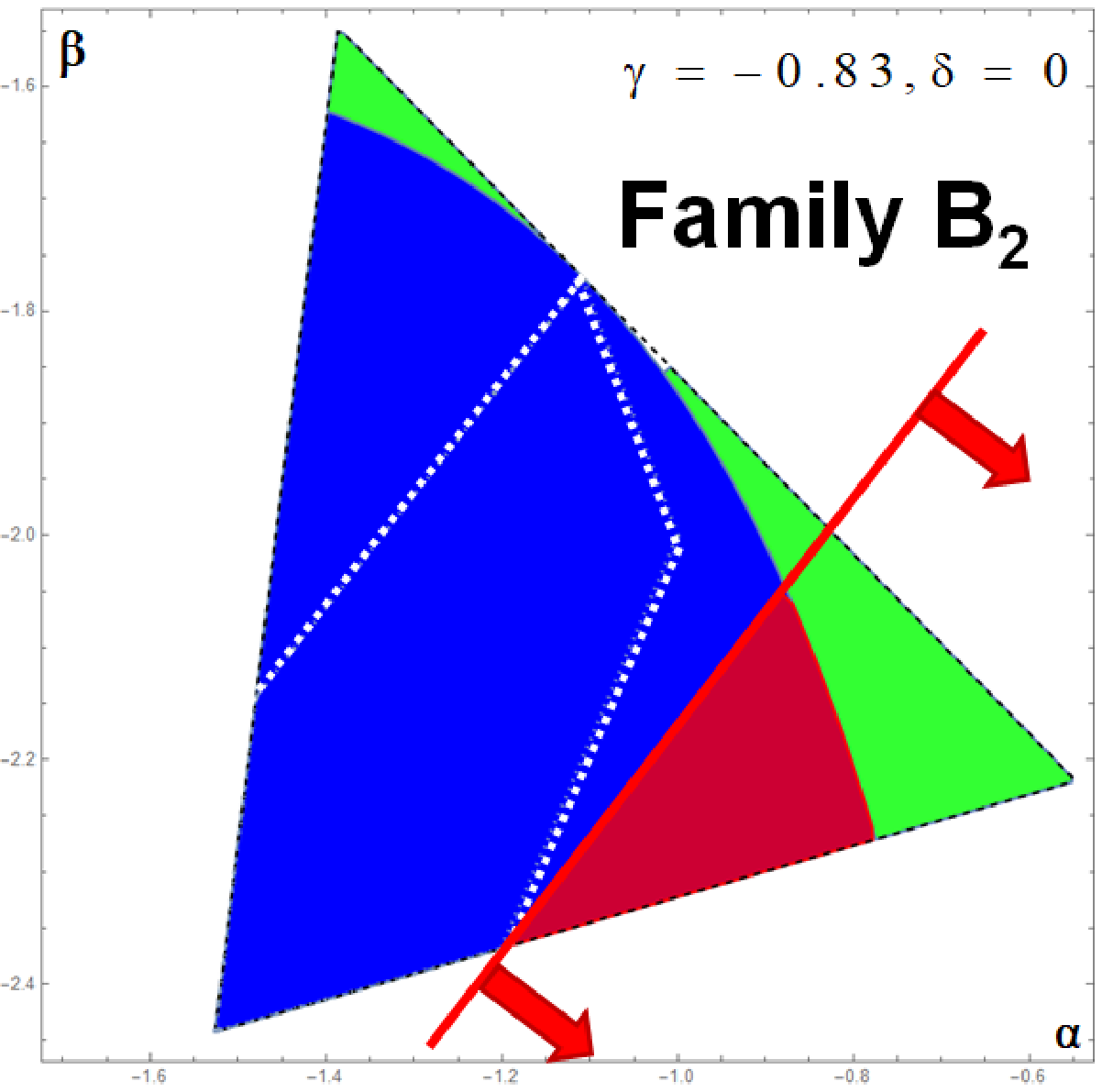}
(d)\includegraphics[width=3.7cm,keepaspectratio=true]{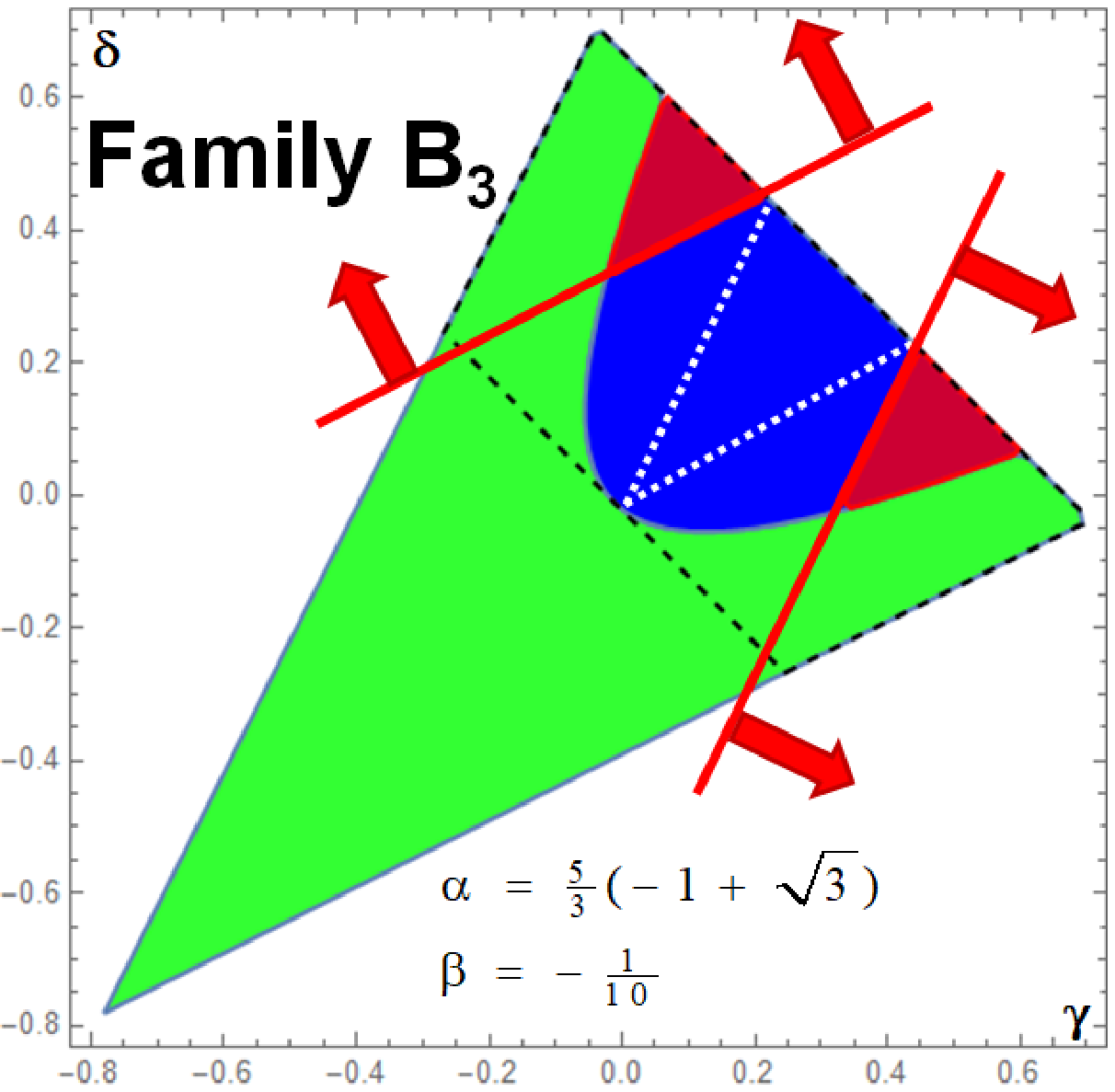}
\caption{These graphics show slices through the $8$--dimensional Magic Simplex: Fig.~(a) shows an example for Family A, where the non-linear entanglement witness (quasipure approximation) is successful but the MUB witness fails. Fig.~(b)-(d) show examples for Family B where the linear MUB witnesses are successful, but the quasipure approximation fails. The colors indicate if the corresponding state is FREE entangled (green), PPT (blue and red) or BOUND entangled (red). The red lines (with arrows) represent one MUB witness and the arrow indicates which states are detected as entangled. The black dashed lines correspond to the enclosure polytope $\mathcal{P}$, whereas the white dotted lines correspond to the kernel polytope $\mathcal{P}$, i.e. states within the white dotted area are certainly SEP.}\label{families}
\end{center}
\end{figure*}

\subsection{Examples: two families}

For illustration of the very problem and how ML algorithms can tackle the problem, let us define two families, particular slices via the magic simplex, which we study by means of analytical entanglement witnesses. As we show here entanglement witnesses that are successful for the first family $A$ fail for family $B$ and vice versa, illustrating the very crux of entanglement detection. In the following we use those families for a check of the reliability of the ML algorithms.

Family $A$, is defined by the mixture of $3$ Bell states that form a line in the simplex with the maximally mixed state, i.e. a chosen Bell state and a chosen Weyl operator to generate the remaining two Bell states. This state has a high symmetry and therefore with a couple of developed tools~\cite{simplex1,quasipure} a witness has been constructed that detects all bound entangled states in this family. Thus the separability versus free/bound entanglement problem can be (analytically) fully solved for this family (by the quasispin criterion defined in appendix~\ref{quasispincriterion}). Let us parameterize this family by $\{c_{kl}\}\equiv \vec{c}$
\beq\label{familyA}
\textrm{Family $A$:}
\quad \vec{c}_A&=&\frac{1}{9} \left(
\begin{array}{c}
  8 \alpha -\beta-\gamma +1 \\ \-\alpha +8 \beta-\gamma +1 \\ -\alpha
   -\beta +8 \gamma+1 \\
 -\alpha -\beta-\gamma +1 \\ -\alpha -\beta-\gamma +1 \\ -\alpha
   -\beta-\gamma +1 \\
 -\alpha -\beta-\gamma +1 \\ -\alpha -\beta-\gamma +1 \\ -\alpha
   -\beta -\gamma+1 \\
\end{array}
\right)\;,
\qquad
\textrm{Family $B$:}\quad \vec{c}_B\;=\;\frac{1}{9}
\left(
\begin{array}{c}
 \frac{8 \alpha }{5}-\frac{\beta }{4}-\gamma -\delta +1 \\ -\frac{\alpha }{5}+\frac{7 \beta }{8}-\gamma -\delta +1 \\ -\frac{\alpha
   }{5}+\frac{7 \beta }{8}-\gamma -\delta +1 \\
 -\frac{\alpha }{5}-\frac{\beta }{4}+2 \gamma -\delta +1 \\ -\frac{\alpha }{5}-\frac{\beta }{4}+2 \gamma -\delta +1 \\-\frac{\alpha
   }{5}-\frac{\beta }{4}+2 \gamma -\delta +1 \\
 -\frac{\alpha }{5}-\frac{\beta }{4}-\gamma +2 \delta +1 \\ -\frac{\alpha }{5}-\frac{\beta }{4}-\gamma +2 \delta +1 \\ -\frac{\alpha
   }{5}-\frac{\beta }{4}-\gamma +2 \delta +1\\
\end{array}
\right)\;.
\eeq
For all those vectors we can label $\vec{c}_A$ by $\{SEP,BOUND,FREE\}$, which is visualized in Fig.~\ref{families}~(a) for $\gamma=0$.  It was the first state family for which a volume of bound entanglement was found and as the picture already suggests the volume of bound entangled states seems very small compared to the volume of free entangled and separable states. In the next section we provide a more quantitative statement underpinning this statement. Let us remark here, it is important to distinguish between the symmetry in the Hilbert space ($8$-dimensional simplex) and the geometry ($9$-dimensional sphere) of the normalized vectors $\vec{c}$.

Totally different entanglement detection methods have shown to be successful in detecting bound entanglement for the family $B$ (defined in appendix~\ref{MUBcriterion}), however,
for this family we do not know everything by means of analytical tools, i.e. we have the labelling $\{FREE,BOUND,SEP,UNKNOWN\}$. Fig.~\ref{families}~(b), ($B_1:\gamma=-\frac{1}{\sqrt{3}},\delta=0$), illustrates the region of the greatest value violating the upper bound on a MUB-witness (defined in the appendix~\ref{MUBcriterion}). This witness is from the experimental point of view very attractive since it contains a simple recipe of how to realize it, i.e. it needs only projections onto MUB-vectors of Alice and Bob, where a particular combination reveals entanglement. Therefore, this particular family triggered an experiment for two photons entangled in their orbital angular momentum ($3$ degrees of freedom) and therefore provided the first proof for an experimental detection of the bound entangled states~\cite{Hiesmayrboundexperiment}. The next slice,  Fig.~\ref{families}~(c) ($B_2:\gamma=0.83,\delta=0$), shows a slice inside $\mathcal{P}$ for which also a MUB-witness not based on a complete MUB set successfully detects bound entanglement. Thus fewer measurement settings are enough to detect bound entanglement. Last but not least, Fig.~\ref{families}~(d), ($B_3:\alpha=\frac{5}{3}(-1+\sqrt{3}),\beta=-\frac{1}{10}$), shows a slice where two different MUB witnesses detect bound entangled states (a third one not depicted is equivalent to the polytope boarder, showing that the symmetry has be carefully taken into account.

\subsection{Detecting Entanglement Within the Magic Simplex: Analytical Methods}

Since we know that all states outside of the enclosure polytope $\mathcal{P}$ are free entangled we only need to consider states within this polytope. The cyclic subgroups within the simplex are further crucial ingredients in exploiting entanglement since having found one state implies that one has found it for $4$ states in the case of $d=2$ and for $12$ states in the case of $d=3$. This is geometrically obvious for $d=2$ since each vertex has the same geometry with respect to the origin ($=$ maximally mixed state), see Fig.~\ref{simplexqubits}.

Free entanglement is efficiently detected via the PPT criterion, to test whether a simplex states belongs to the kernel polytope $\mathcal{K}$, and thus is certainly separable, needs to solve the linear equations defining $\mathcal{K}$ with $12$ unknown variables, which is quite time consuming and needs a careful programming. In Fig.~\ref{families} all points failing the PPT criterion are colored ``green'' and those passing the kernel test are within the white dotted area. BOUND entangled states can be detected by suitable entanglement witnesses, i.e. Hermitian observables, where we applied two different ones (defined in appendix~\ref{quasispincriterion} and~\ref{MUBcriterion}), those states are colored ``red''. Note that the first criterion is a non-linear witness, whereas the MUB-witness is a linear one. Note also that e.g. if bound entanglement is detected, not necessarily all free entanglement is automatically detected.

\section{Solution by the Discretized Magic Simplex: the ground truth}

Now we proceed by discretizing the simplex to compute for each vector the label FREE, BOUND, SEP or UNKNOWN if any of our above described methods fail. In the case of $d=2$ the simplex fills a volume of $\frac{1}{6}$ ($17\%$) of the enclosing cube. And the enclosure polytope equals the kernel polytope, $\mathcal{P}=\mathcal{K}$, and fills half of the volume. This is unfortunately very different to the case of $d=3$. The simplex fills only about $0.009\%$ of the cube and the polytope about $58\%$ of the simplex, which means that without the need of checking the PPT criterion $42\%$ are certainly free entangled. Consequently, for the generation of data vectors for the machine learning algorithms we can concentrate on data vectors that are within the polytope $\mathcal{P}$ (in Fig.~\ref{families} depicted by a dashed (black) lines).

\subsection{Reducing to the Enclosure Polytope $\mathcal{P}$}

As for most of the cases the ground truth, i.e. the generation of the data (or its selection) is crucial for the success of any machine learning algorithm and it is usually the most time consumption process, which was here the case and was performed with the help of a supercomputer. The first crucial questions is the stepsize $\Delta$, if it is too big none or too less bound entangled states are detected, if it is too small the computation times exceeds to several months.

Therefore, our general strategy is to generate vectors $\vec{c}$ where $8$ out of $9$ components are varied in the interval $[0,\frac{1}{3}]$ with a certain stepsize $\Delta$ (the $9th$ component is given by the normalization $|\vec{c}|=1$). Checking of each point $(\frac{1}{\Delta}+1)^8$ the positivity results in the set of magic states within the enclosure polytope $\mathcal{P}$. The next step is to check for those vectors the $PPT$--criterion (defined in the appendix~\ref{pptcriterion}), thus to separate the set into those that are free entangled (labelled by FREE) or not. If the state was not labelled FREE, we performed the kernel test (if passing labelled by SEP), which is unfortunately not always reliable since the programm has to check for solutions of a system of equations which depends strongly on the input form. Here we had to post-check data vectors. The states failing the kernel test could be either BOUND entangled if one of the entanglement witness tests (analytical and numerical tests described later) passed or as UNKNOWN if not.

Thus in summary we generated magic simplex states within the enclosure polytope $\mathcal{P}$ which are labelled by FREE, SEP, BOUND or UNKNOWN. One strategy is to remove the UNKNOWN states and let the machine learning algorithms learn from the remaining states or if one is confident to have found close to all bound entangled states, to label the UNKNOWN states as SEP states. Labeling the opposite (UNKNOWN$\rightarrow$ BOUND) and comparing the machine learning solutions over consistency checks may give a check whether all bound entangled states where found. Generally, to characterize the performance of the machine learning besides the typical parameters as accuracy, the confusion matrix plot, the feature scores,\dots, we apply them to the examples states (Family A, Family B), see Fig.~\ref{families}.

In the following two sections we present the solution of the above procedure for a sub-state space in the magic simplex, for which we know everything and then proceed to the complete magic simplex state for $d=3$.

\subsection{Solution for Family A}

Let us consider all states of Family A, e.g. explicitly given by (note $\sum P_{k,l}=\mathbbm{1}$)
\beq
\rho_A&=&\frac{1-\alpha-\beta-\gamma}{9}\, \mathbbm{1}_9+\alpha\; P_{00}+\beta\; P_{0,1}+\gamma\; P_{0,2}\qquad\Leftrightarrow\qquad \vec{c}_A\;.
\eeq
Since all Bell states lie on a line, this exhibits a high symmetry such that an optimized witness can be guessed and, subsequently, its optimality in detection all existing bound entanglement could be proven~\cite{simplex1,simplex3}. Later on it was found that this non-linear witness is equivalent to the quasi-pure approximation (see appendix~\ref{quasispincriterion}). This result presented the first volume of bound entangled states in the Hilbert space and opened the possibility of a potential experimental observation, however the volume was found to be too small as can be directly deduced from Fig.\ref{families}~(a) by eye. Moreover, it is not clear, how this non-linear entanglement witness should be experimentally realised. This is very different to linear MUB witnesses that contain a rather simple protocol of how to realize them experimentally.

To quantify the volume of bound entangled states within the $3$--dimensional simplex spanned by any three Bell states on a line, we vary the three parameters $\alpha,\beta,\gamma$ by a stepsize $\Delta=\frac{1}{72}$, i.e. testing $614.125$ vectors for which we find $95.455~(15\%)$ satisfying the positivity condition (spanning a $3$--dimensional simplex). Out of those states $78.042~(81.8\%)$ are free entangled, from which $74418 ~(78.0\%)$ are outside of the polytope, thus those are certainly free entangled. From the remaining $18.2\%$ states $17.317~(18.1\%)$ are separable and only $96~(0.1\%)$ are bound entangled.

In summary, only $3.8\%$ within this enclosure polytope $\mathcal{P}_A$ are free. Thus for these $12$ subspaces of the $8$--dimensional simplex the enclosure polytope $\mathcal{P}_A$ is a quite good separability border in contrast to the kernel polytope $\mathcal{K}_A$. This is in strong contrast to the result of the full magic simplex that we present in the following.

\subsection{The full magic simplex (based on analytical witnesses)}

Finally, we were able to scan the enclosure polytope $\mathcal{P}$, which covers about $58\%$ of the total simplex, by a stepsize of $\frac{1}{18}$, big enough to detect bound entangled states, but small enough to limit the computing time on a supercomputer two weeks in total. This corresponds to $7^8=5. 764. 801$ vectors that had to be checked for positivity, resulting in a subset of $899.857~(15.6\%)$ magic simplex states. Those states we checked via the PPT criterion allowing for labelling $620.406~(69\%)$ as FREE entangled states within $\mathcal{P}$, which is in strong contrast to the subspace solution of Family A, i.e. less than $\sim\frac{1}{3}$ states are either separable or bound entangled.

For the remaining states we check for whether they belong to the kernel $\mathcal{K}$ which is not computationally simple since a system of $12$ equations have to be tested for a possible solution. We found only $27.055~(3\%)$ meet the condition for being within $\mathcal{K}$. For the remaining states ($18\%$) we checked for bound entanglement via two different entanglement witnesses classes. The quasipure criterion detects $37\%$  of all BOUND states,
 the MUB criteria $2\%$ of all BOUND states 
and $0.9\% $ are detected by both criteria. This means that as much as $24\%$ of the data vectors are UNKNOWN exploiting those known sets of witnesses. This renders the problem as not very suitable for a successful application via ML algorithms. 

\subsection{The full magic simplex (based on additional numerically generated witnesses)}

Therefore, we exploited the recently discovered fact that an entanglement witness has an upper bound (the very definition of a witness) and in general often also a non-trivial lower bound~\cite{MUB3}. Furthermore, we exploited the structure of the magic simplex by noting that an effective witness in the magic simplex can have the form $W=\sum_{k,l}\kappa_{k,l}\; P_{k,l}$. We generated over $9000$ witnesses by random number $\kappa_{k,l}$ and computed via the help of a proper parametrization of unitaries~\cite{Hiesmayrunitaries} the upper and lower bounds given by the optimum over all separable states: $\min_{\rho_\textrm{SEP}}Tr(W_i\rho_{SEP})\leq Tr(W_i\rho)\leq \max_{\rho_\textrm{SEP}}Tr(W_i\rho_{SEP})$. Those witnesses we applied to the remaining set generating new witnesses until no further bound entangled state was detected via this method. This is a very time consuming process and was undertaken by the Vienna super-computers. By this method we could detect further $142.011$ states to be bound entangled.

In summary, via this procedure we find $70\%$ to be free entangled, as much as $17\%$ to be bound entangled and $3\%$ to be certainly separable in $\mathcal{P}$ (see also Fig.~\ref{simpelexsummary}). Thus $10\%$ remain as UNKNOWN. Assuming that we found all bound entangled states, those states should be separable due to the convexity argument, namely that the set of separable states is convex. As we show in the following we can test whether those UNKNOWN vectors are more consistent with being SEP or BOUND via ML algorithms. Let us emphasize that irrespective of the labelling of the remaining states, we find surprisingly more bound entangled states than separable states which was not to be expected and shows for the first time that bound entangled states are not negligible (only difficult to detect).

\section{Machine Learning}

In this section we utilize machine learning algorithms~\cite{ML1,ML2} for different purposes. Firstly, we investigate the question of whether the remaining states labelled as UNKNOWN are more likely to be separable or entangled. Secondly, we investigate the question how good machine learning can learn from the data by testing its prediction for some slices in the magic simplex. Thirdly, we ask whether ML can reveal novel unknown structures.

\subsection{Are the remaining UNKNOWN states SEP or BOUND?}

\begin{figure*}
\begin{center}
(a)\includegraphics[width=3.5cm,keepaspectratio=true]{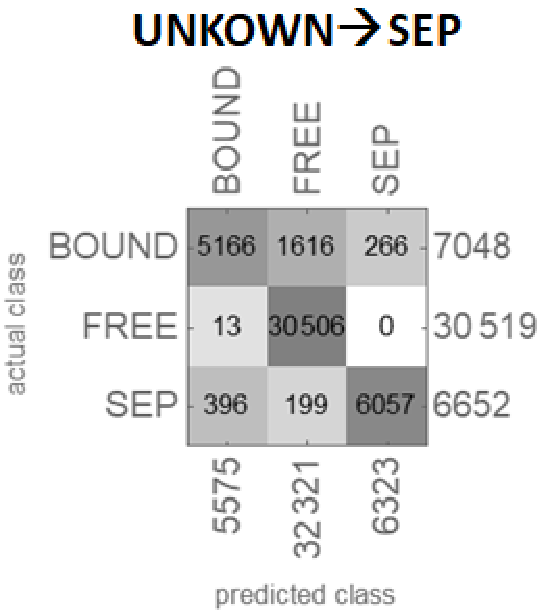}(b)\includegraphics[width=3.5cm,keepaspectratio=true]{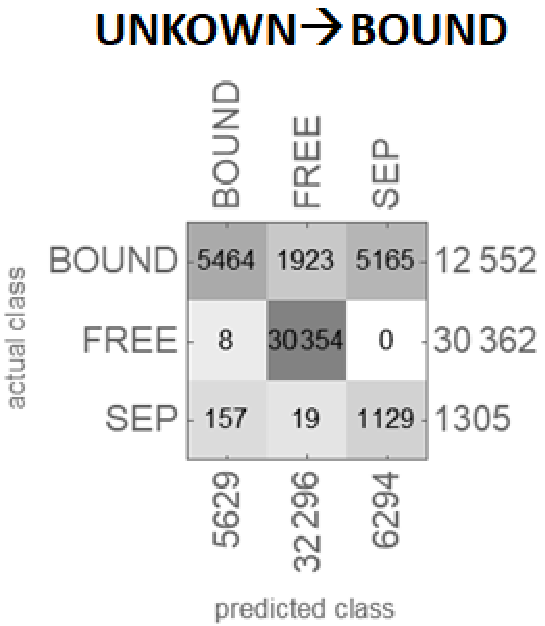}
(c)\includegraphics[width=3.5cm,keepaspectratio=true]{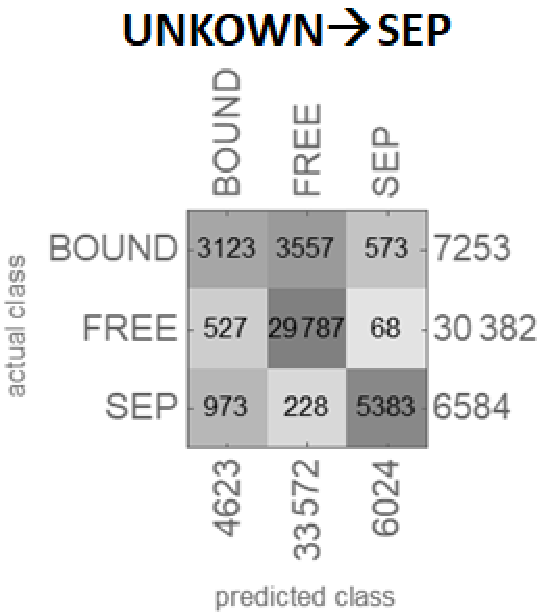}(d)\includegraphics[width=3.5cm,keepaspectratio=true]{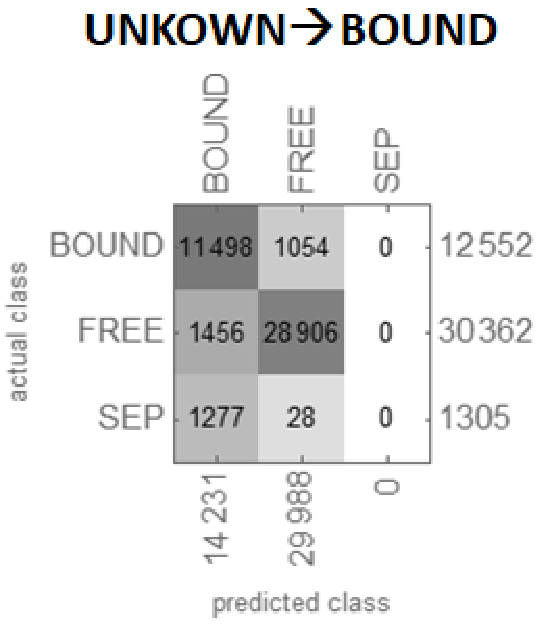}
\caption{Here the matrix confusion plots are presented for the two scenarios, UNKNOWN $\rightarrow$ SEP and UNKNOWN $\rightarrow$ BOUND. (a),(c) present the result for the ``nearest neighbour'' and the ``gradient boosted trees'' ML algorithm for the first scenario and (b),(d) for the second scenario. One observes that for both ML algorithms the second unexpected scenario the SEP labelled points are mostly confused with the BOUND labelled points, but not with FREE, i.e. those points are in average closer to the the set of BOUND than FREE.}\label{confusionplots}
\end{center}
\end{figure*}

Since we know that the set of separable states has to form a convex set in the Hilbert space, we assume that any classification algorithm based on a particular distance definition would learn exactly this feature. For this purpose we exploited a ``nearest neighbor'' algorithm, classifying a class by $k$-nearest neighbours, first for labelling all UNKNOWN as SEP ($15\%$) and secondly as BOUND (only $3\%$ (kernel) are now SEP). We expect, if our witnesses have detected all bound entangled states, that the first case should give a better solution in terms of fitting parameters as we know that the set of separable states is convex and there are more separable states expected than those in the kernel $\mathcal{K}$. Indeed, the accuracy for the first scenario is $94\%$ with probabilities $97\%/82\%/93\%$ of detecting FREE/BOUND/SEP, whereas the second scenario result in a quite lower accuracy of $84\%$ and the probabilities $97\%/60\%/30\%$, i.e. no change for FREE but a significant drop in the detection probability of the two other classes. This signature becomes even stronger for another ML algorithms ``gradient boosted trees'', which is a machine learning technique for regression and classification problems that produces a prediction model in the form of an ensemble of trees, which are trained sequentially with the goal of compensating the weaknesses of previous trees. The confusion-matrix plots are presented in Fig.~\ref{confusionplots}. Those suggests that the first scenario is the more consistent one with the assumption outlined above. In the following we assumed that the UNKNOWN states are indeed SEP.

\subsection{ML Algorithms for SEP versus BOUND (without FREE)}

\begin{figure*}[h!]
\begin{center}
(a)\includegraphics[width=3cm,keepaspectratio=true]{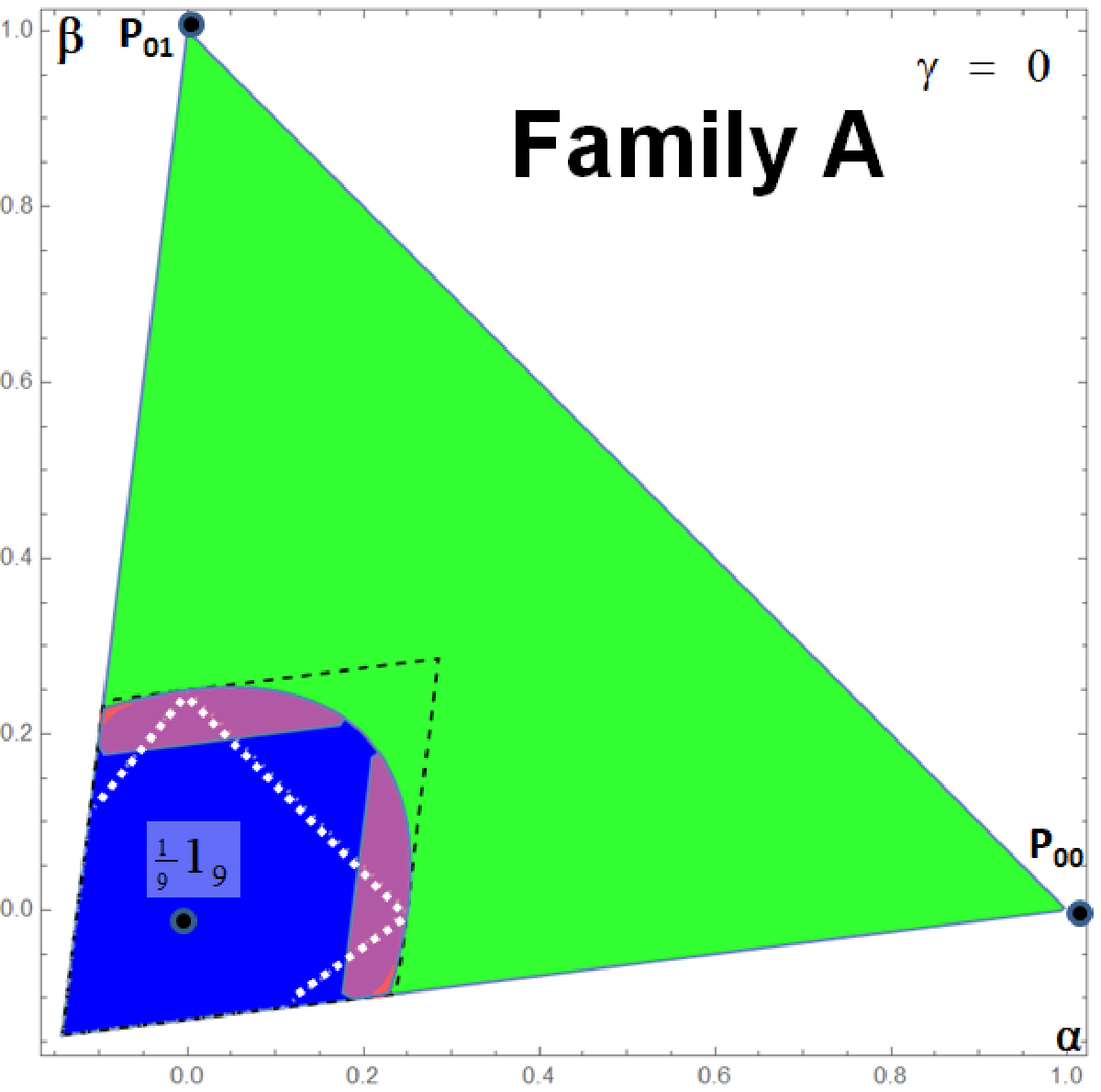}
(b)\includegraphics[width=3cm,keepaspectratio=true]{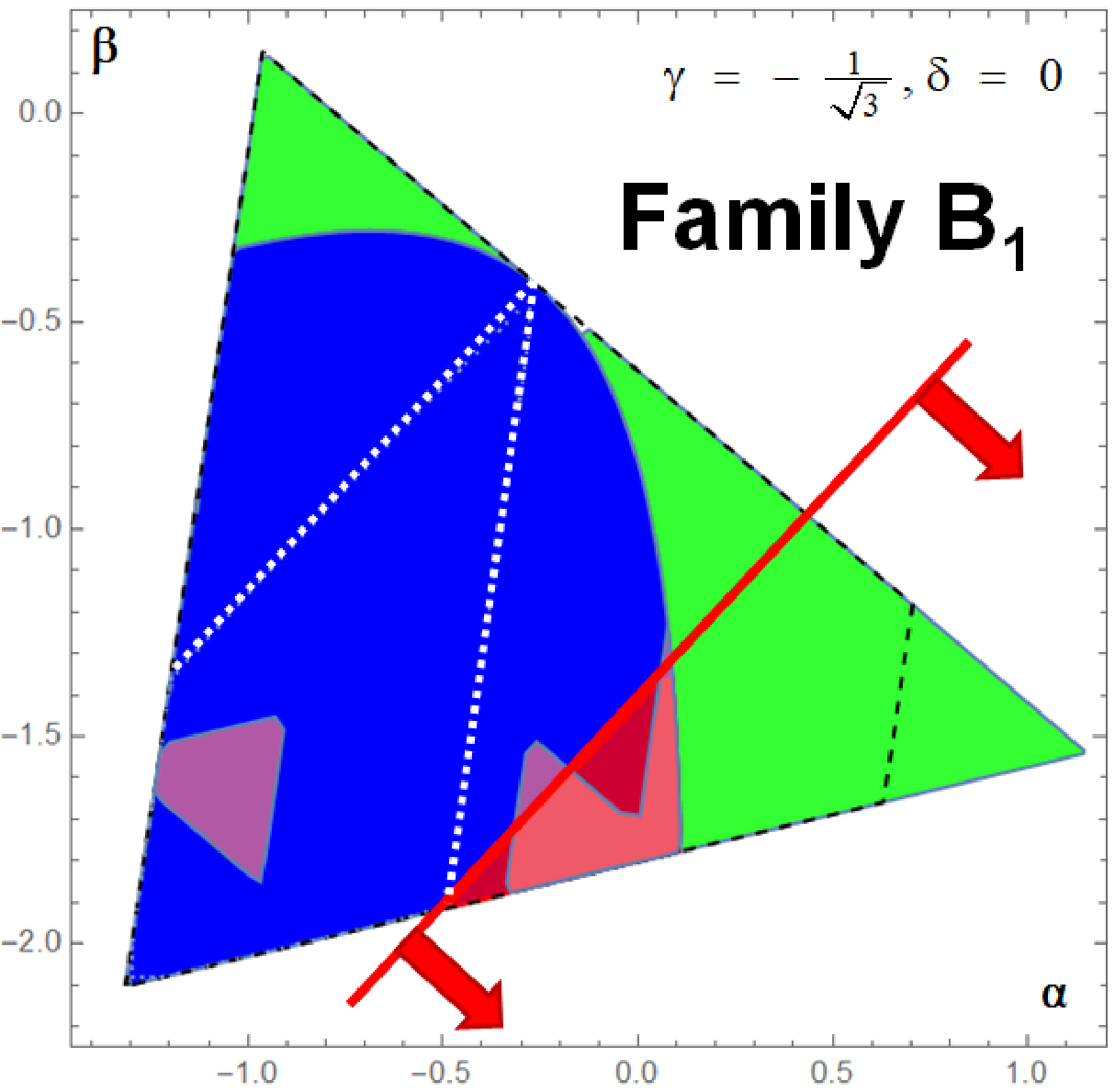}
(c)\includegraphics[width=3cm,keepaspectratio=true]{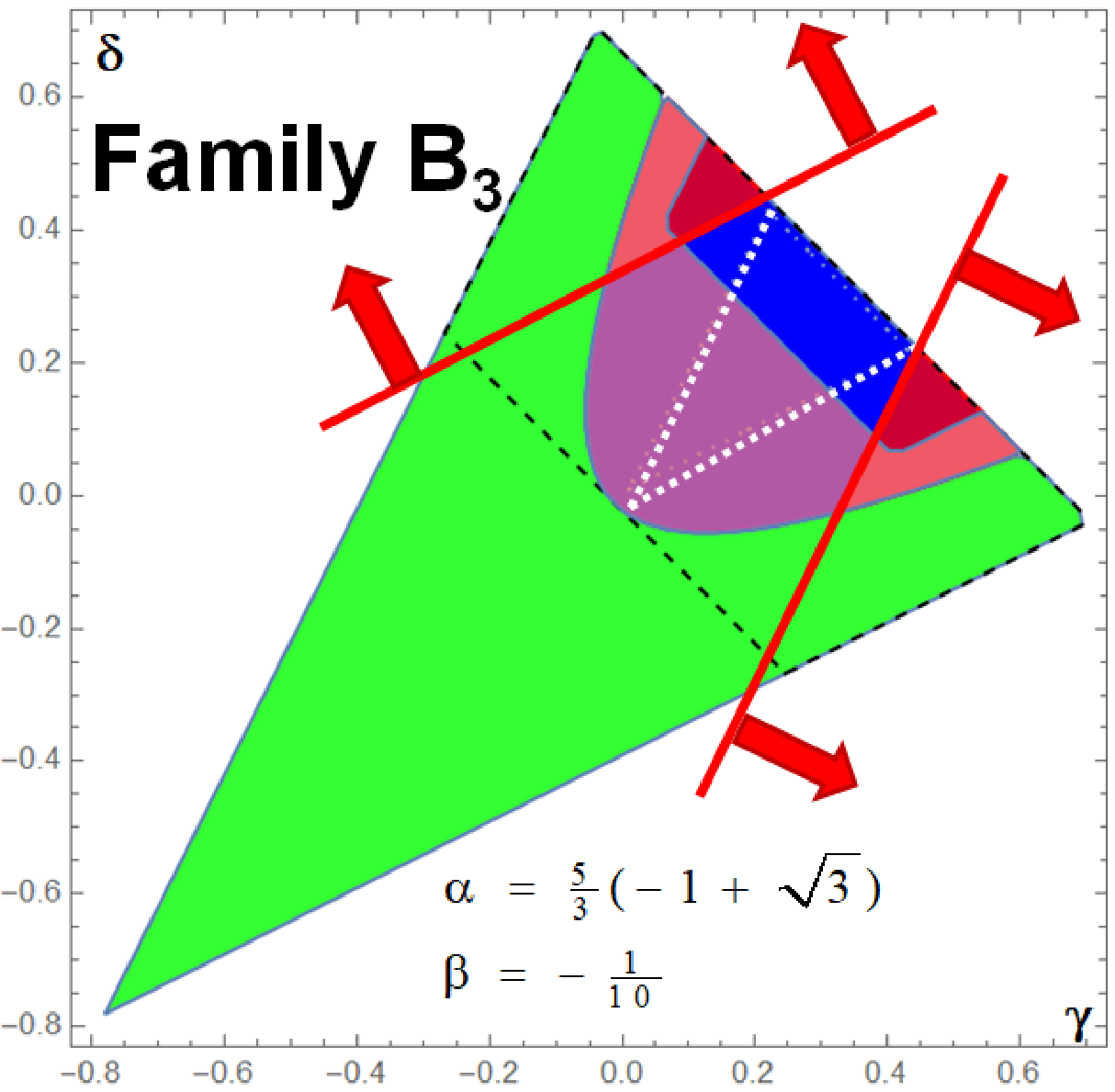}
(d)\includegraphics[width=3cm,keepaspectratio=true]{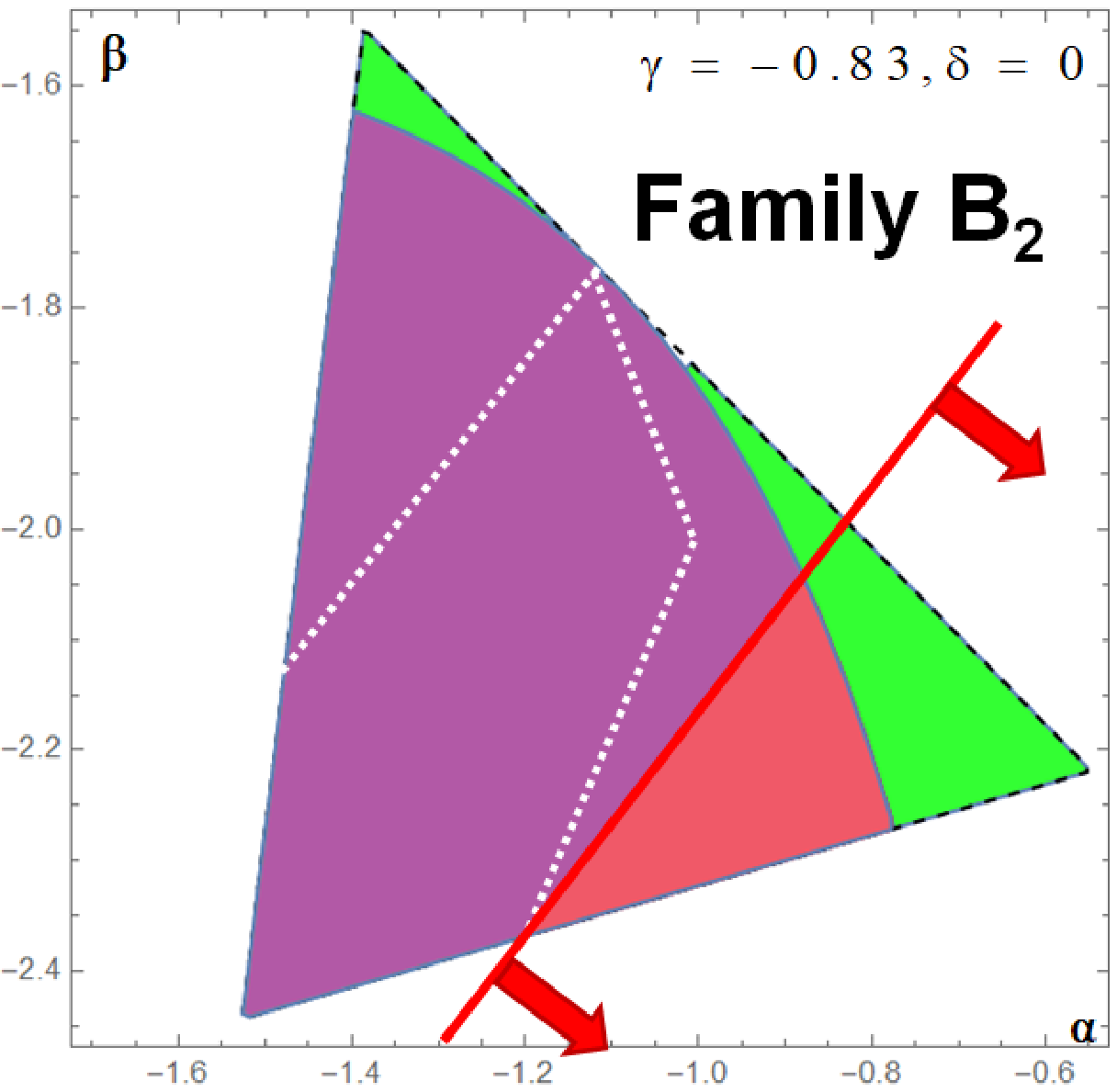}
(e)\includegraphics[width=3cm,keepaspectratio=true]{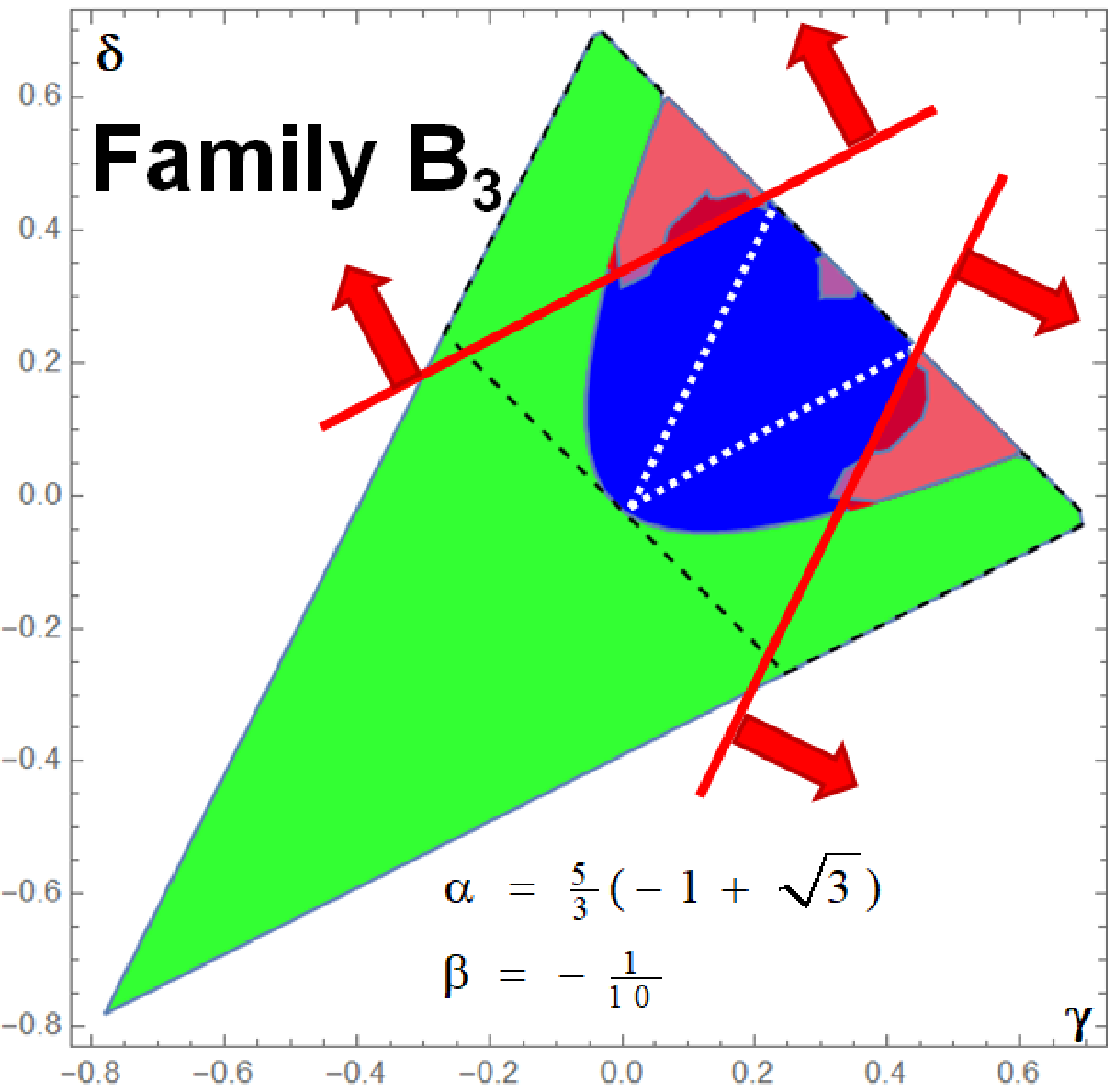}
\caption{These graphics show slices through the $8$-dimensional magic simplex including results from two ML algorithms based on PPT states: ``Random Forest'' (accuracy $91.51\pm0.13\%$, SEP score: $90\%$; BOUND score $92\%$) and ``Nearest Neighbours'' (accuracy $90.64\pm0.14\%$, SEP score: $90\%$; BOUND score $90\%$) . Color coding is as in Fig.~\ref{families}. The light pink area shows the region found by the ML algorithm as BOUND by the majority vote over all $12$ possible realizations, consequently the remaining PPT region is detected as SEP. Figure (a)-(c) show the results for ``Random Forest'' and (d)-(e) for ``Nearest Neighbours'' algorithm.}\label{MLresult1}
\end{center}
\end{figure*}

Assuming that we found all bound entangled state and guided by the considerations of the last sections, we label all UNKNOWN as SEP. Furthermore, we are only interested in distinguishing SEP from BOUND since free entanglement can be faithfully detected by the PPT criterion if the full knowledge of the physical state is available. Now we train different ML algorithms and apply them to our exemplary slices of family $A$ and family $B$.  Note that those slices are not necessarily the ground truth, i.e. only a few points of the slices may be within the ground truth. Furthermore, we know from geometry considerations that the results for our exemplary slices have to be identical for all $12$ geometrical equivalent representations. This fact was not learnt efficiently by our ML algorithms, therefore to find a definite solution, whether the ML algorithm labels geometrically identical points as SEP or BOUND, we introduce a ``majority vote', i.e. if the label BOUND is more than $5$ the result is labelled as BOUND. Exemplary results are visualized in Fig.~\ref{MLresult1}. Here we show the solution for two ML algorithms (``random forest'',``nearest neighbour'') applied to our four slices in the magic simplex. The accuracy is quite hight ($\approx 90\%$), though both algorithms are either overshooting or undershooting for the families $A,B$. Interestingly, the ML results would suggest that for families $B_2,B_3$ we should expect more bound entanglement than detected by the analytical methods.

\subsection{Dimension Reduction Algorithms}

\begin{figure}[t]
\begin{center}
(a)\includegraphics[width=3.8cm,keepaspectratio=true]{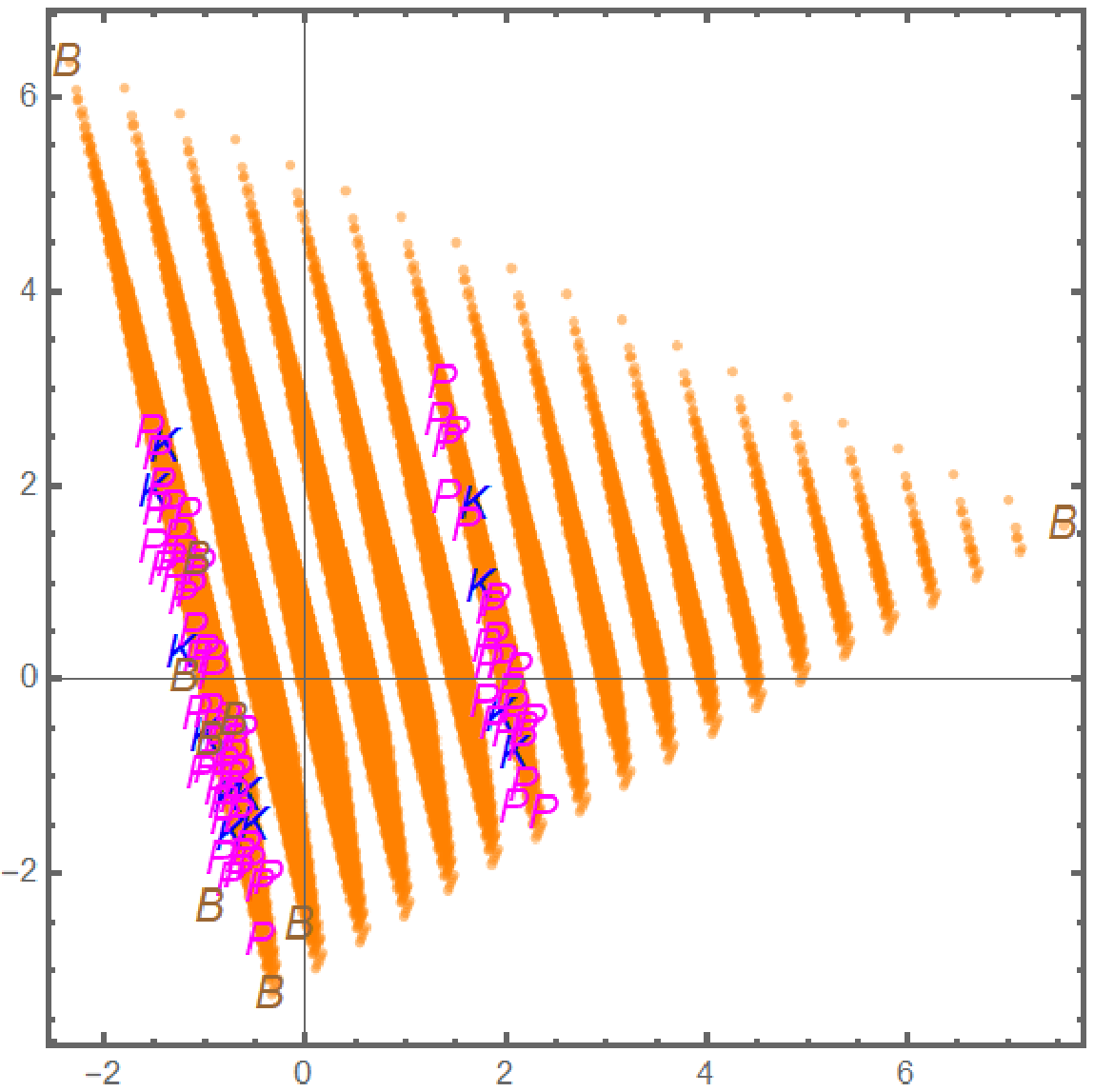}
(b)\includegraphics[width=3.8cm,keepaspectratio=true]{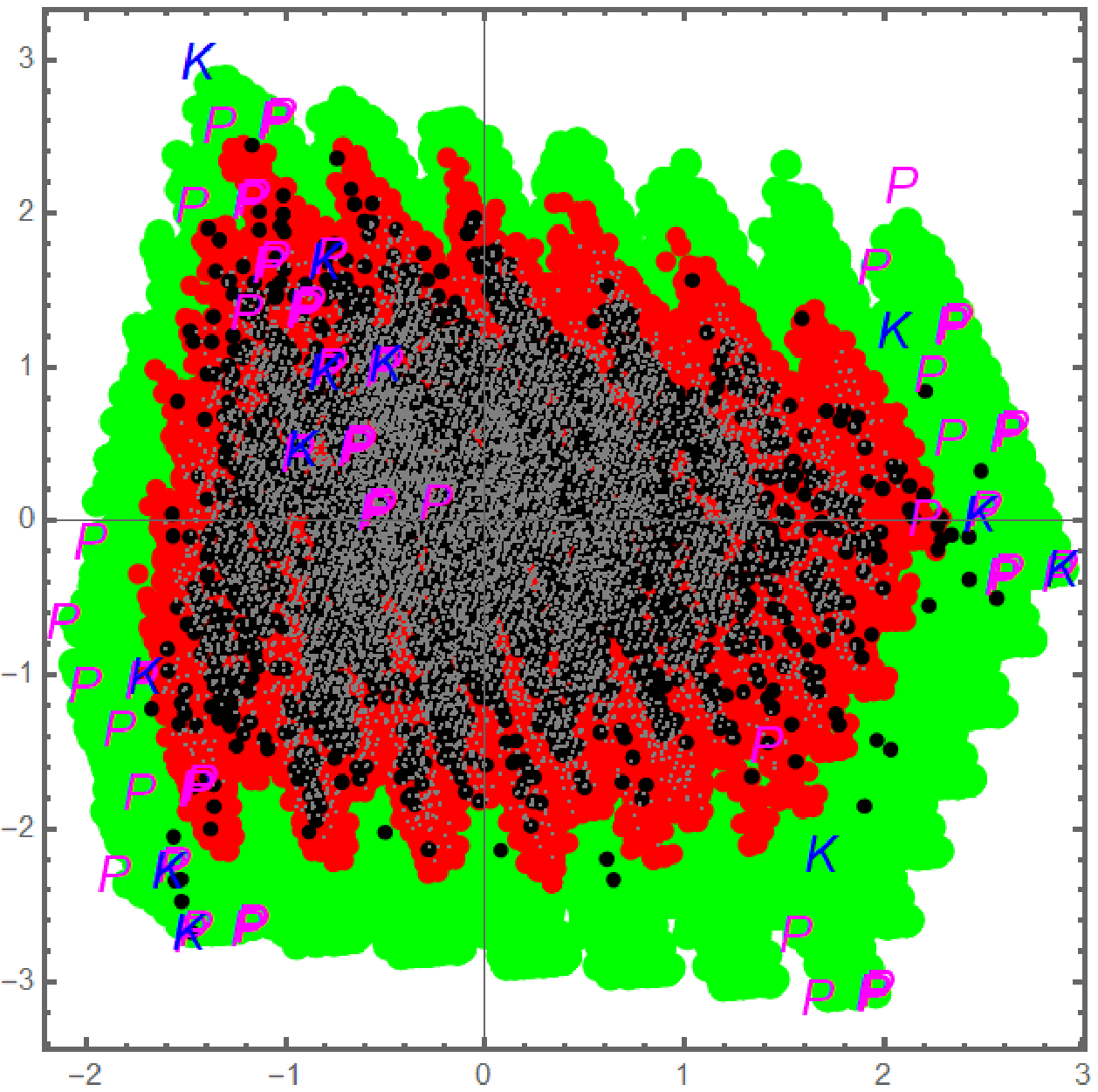}
(c)\includegraphics[width=3.8cm,keepaspectratio=true]{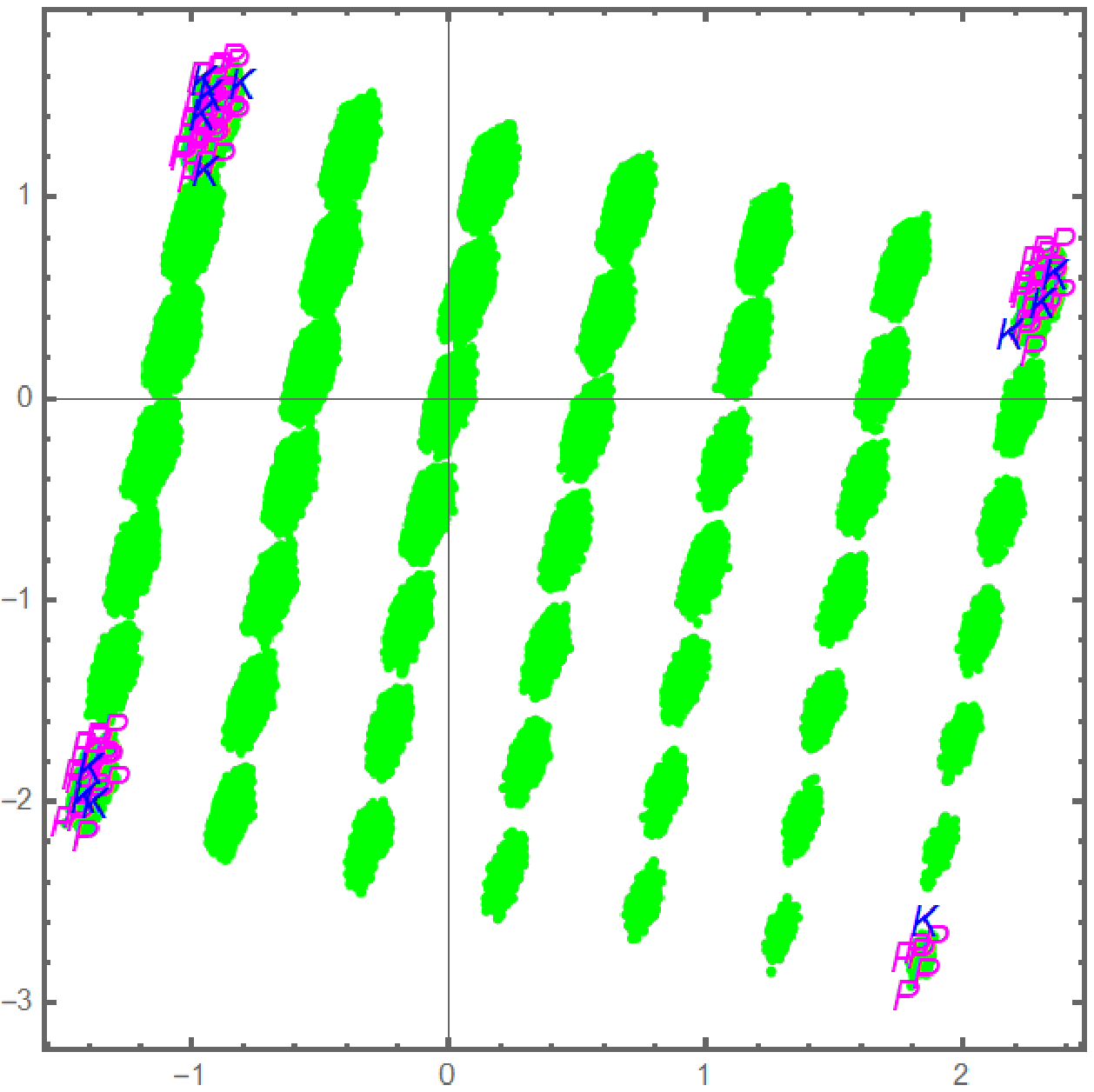}
(d)\includegraphics[width=3.8cm,keepaspectratio=true]{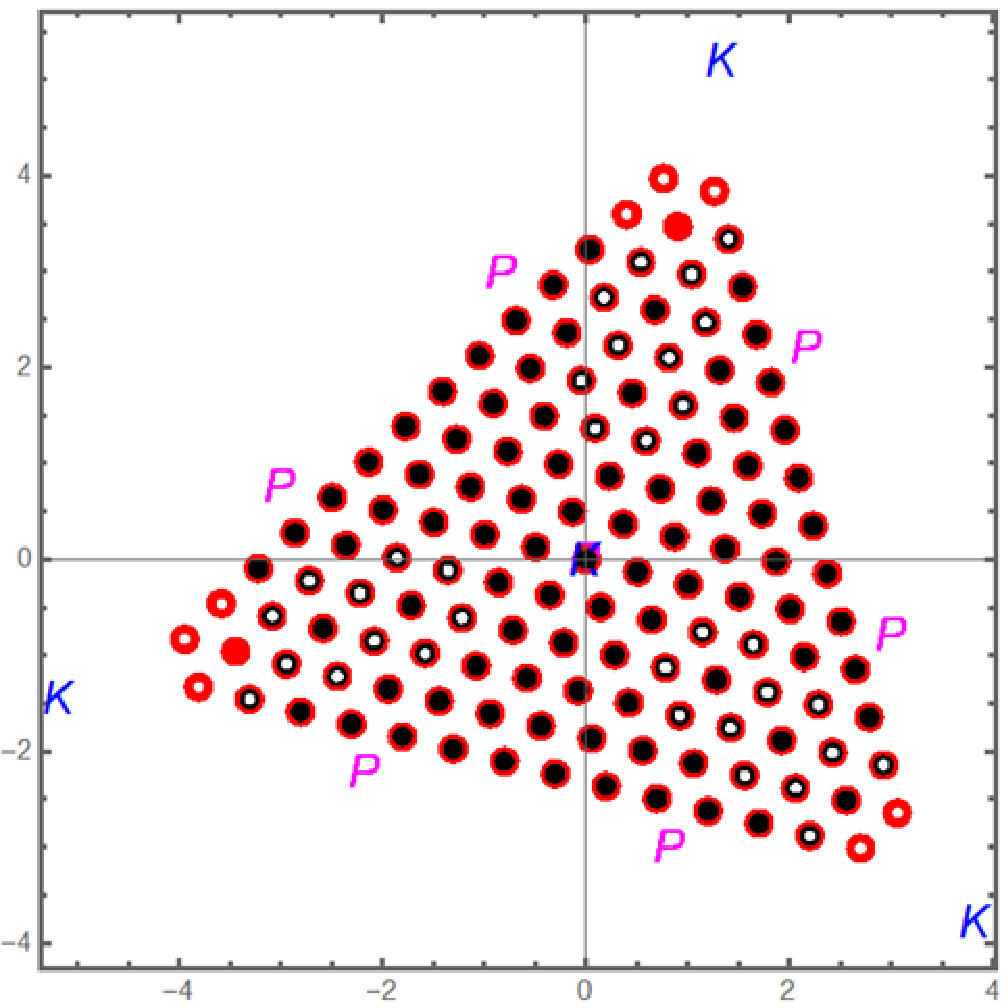}
\caption{These graphics present results of a two-dimensions reduction based on PCA for (a) the full data set of the simplex  and (b) for the polytope $\mathcal{P}$, (c) for the set of FREE vectors, (d) of BOUND vectors. The color coding is FREE=green, BOUND=red, SEP=black and UNKNOWN=grey. The letter $K$ (blue) denotes the results of the kernel polytope $\mathcal{K}$, whereas the $P$'s (magenta) denote the remaining vertices of the enclosure polytope $\mathcal{P}$ and $\mathcal{B}$ denote the nine Bell states. Those points give a good orientation of the resulting projection view, however, note that those are not necessarily included into the data set. In the graphic (d) all bound states reduce to $133$ points, where the states that are red\&black or red\&white correspond to states which are detected by the quasi-spin criterion and the MUB criterion, respectively.}\label{pca1}
\end{center}
\end{figure}

\begin{figure}[t]
\begin{center}
(a)\includegraphics[width=3.8cm,keepaspectratio=true]{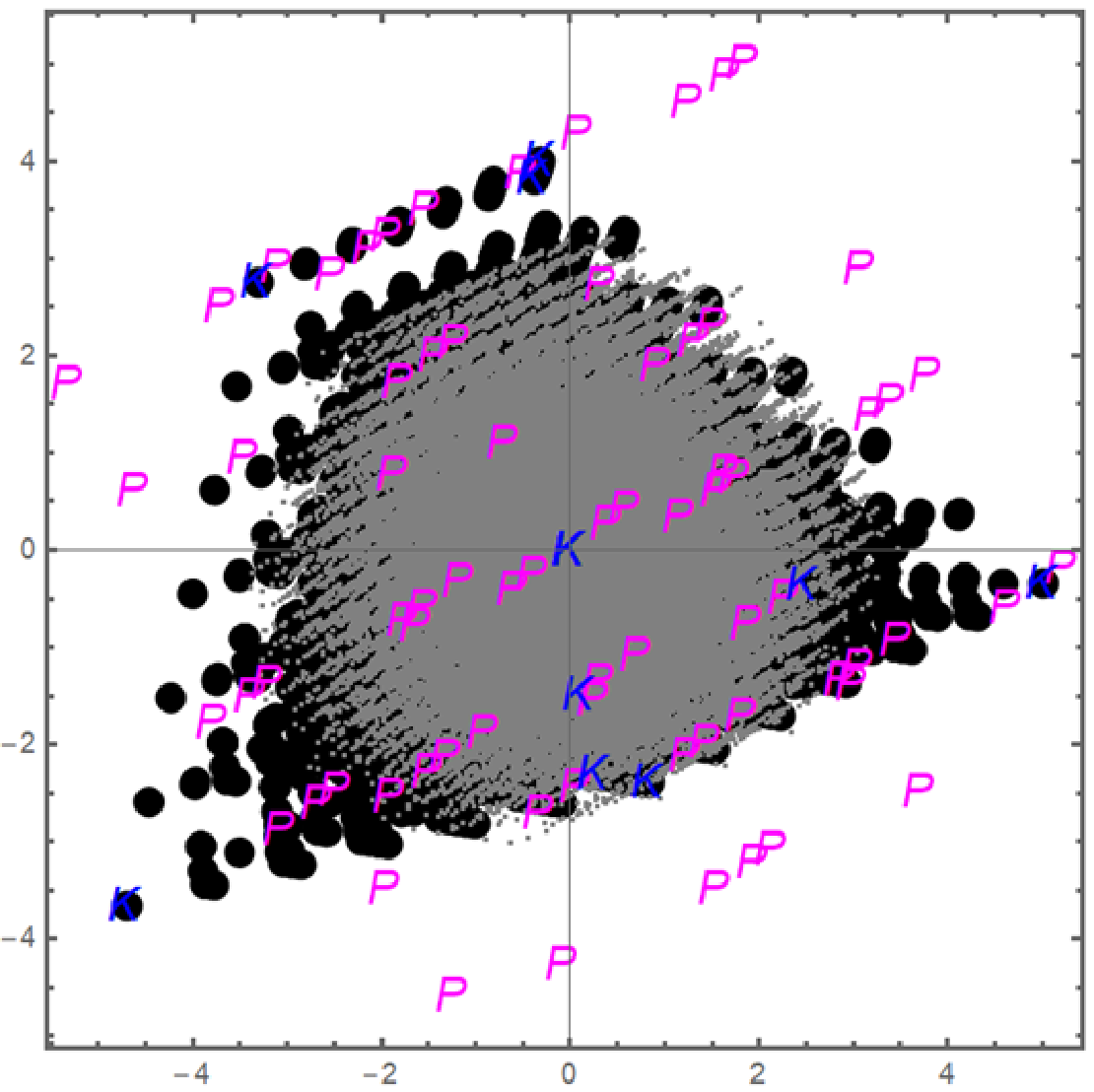}
(b)\includegraphics[width=3.8cm,keepaspectratio=true]{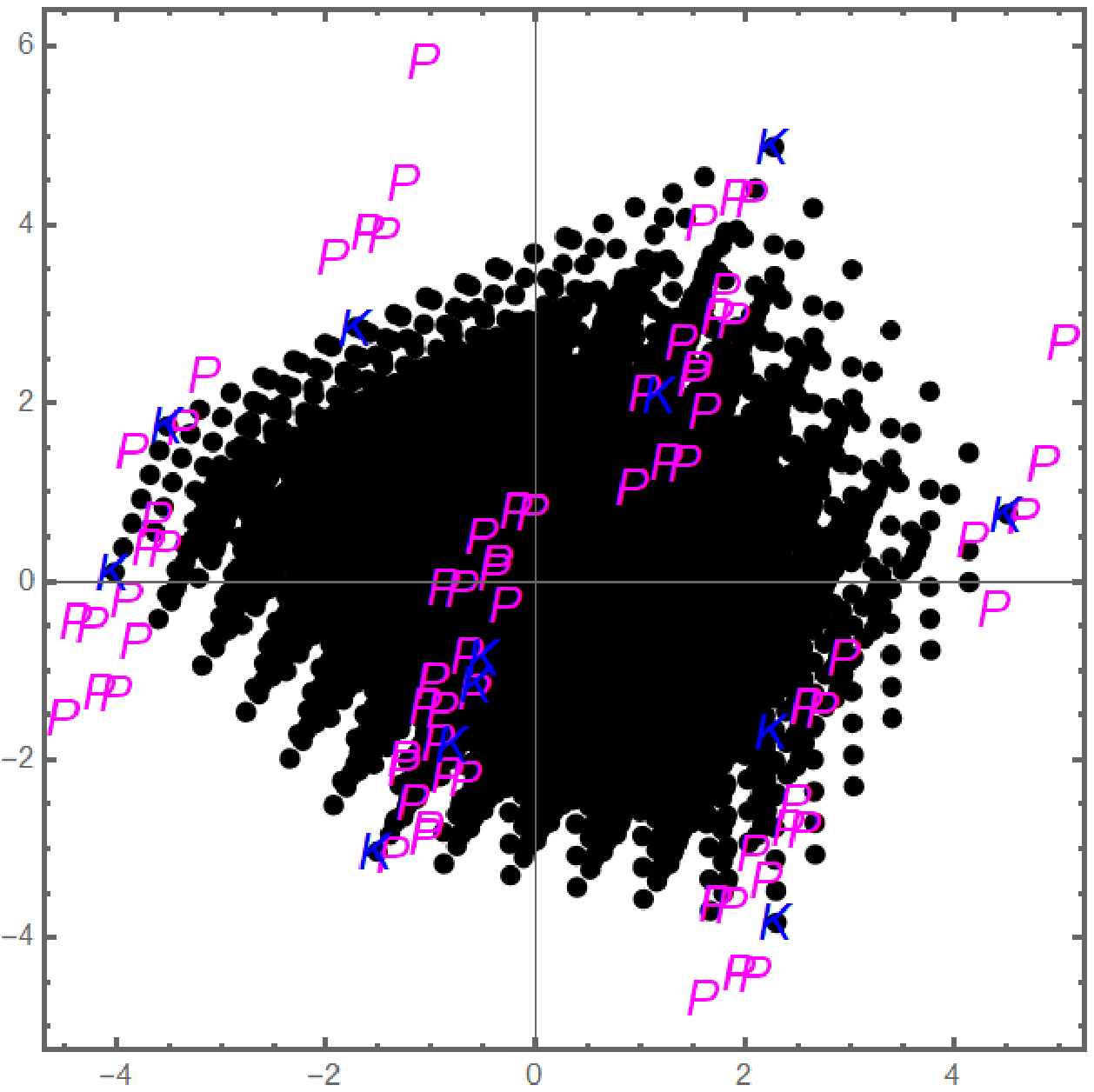}
(c)\includegraphics[width=3.8cm,keepaspectratio=true]{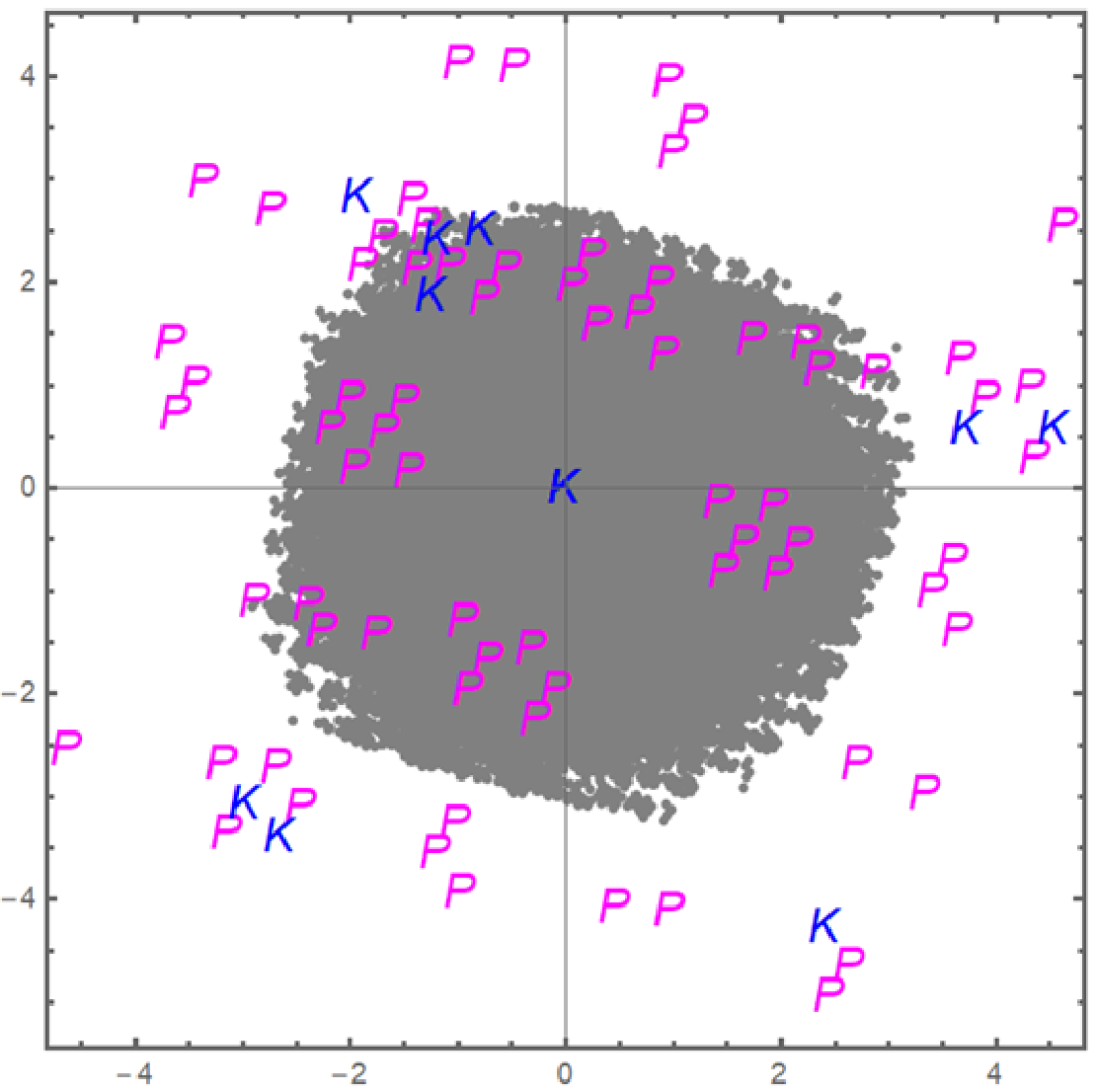}
(d)\includegraphics[width=3.8cm,keepaspectratio=true]{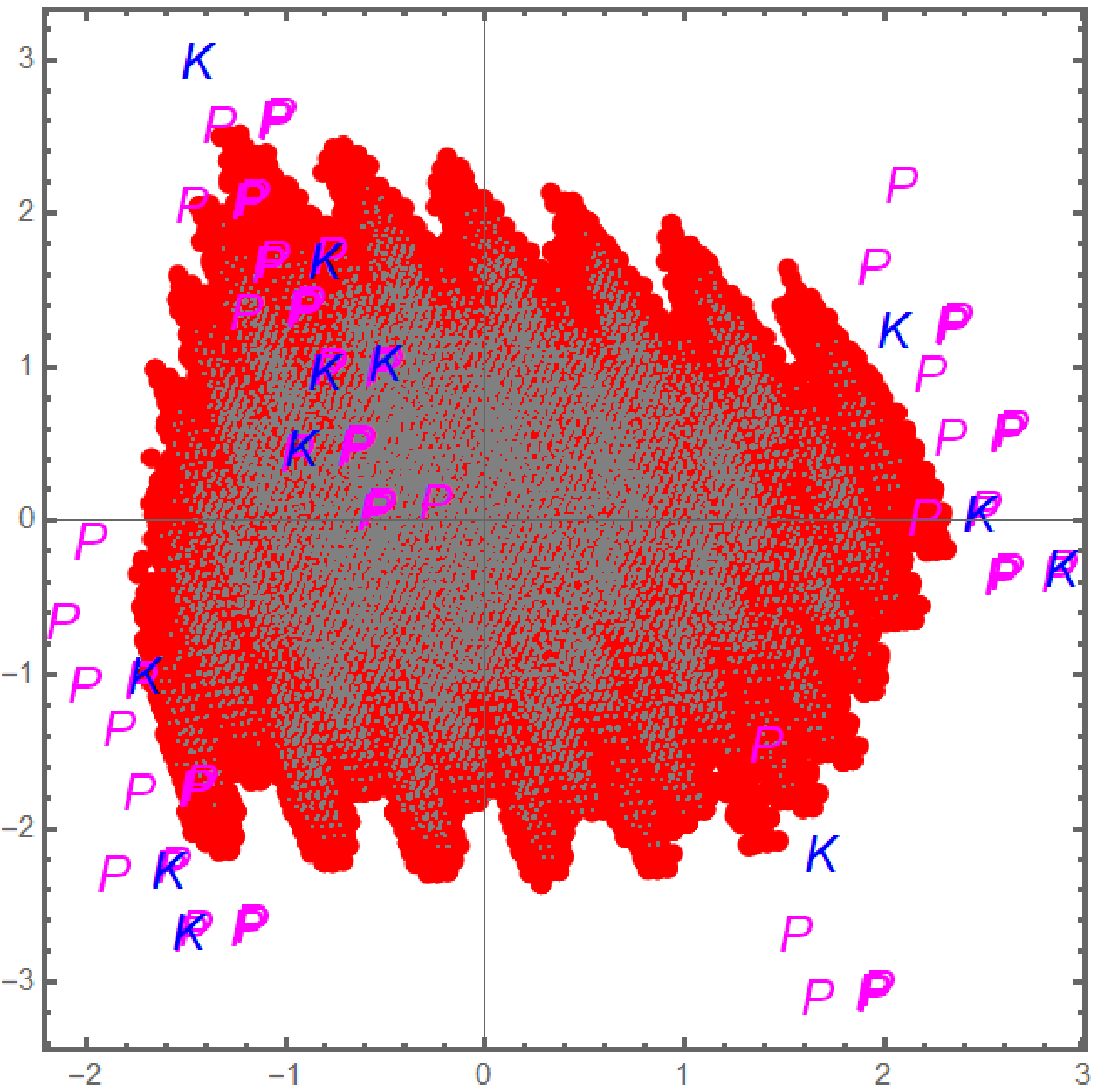}
\caption{These graphics present results of a two-dimensional reduction based on PCA for (a) kernel and UNKNOWN vectors, (b) for the kernel vectors, (b) for the set of UNKNOWN vectors, (c) of BOUND and UNKNOWN vectors. The color coding is FREE=green, BOUND=red, SEP=black, UNKNOWN=black/gey. The letter $K$ denotes the results of the kernel polytope $\mathcal{K}$, whereas the $P$'s denote the remaining vertices of the enclosure polytope $\mathcal{P}$. Those points give a good orientation of the resulting projection view, however, note that those are not necessarily included into the data set.}\label{pca2}
\end{center}
\end{figure}

A standard method in ML is to reduce the dimension and possibly noisy data. We know that our data has in the Hilbert space a strong eminent symmetry connected with $3$, the dimension of the two quantum subsystems. Therefore, applying dimension reduction algorithms seems fruitful since the characteristic features may be compressed in some principal components due to the geometrical constraints (simplex, polytope,$\dots$).

One of the most popular dimension reduction algorithms is based on the PCA (Principal Component Algorithm). It works by taking all data vectors, de-meaning it, and computing all co-variances. From this covariance matrix the eigenvectors and eigenvalues are computed and ordered by the largest one. Note that other dimension reduction algorithms gave similar results for our data, for simplicity we show only the results obtained by the method PCA. Moreover, the results depend on the ordering of the input data, so we tried different approaches which give in total similar results, particularly for the most important set of states, the bound entangled states (Fig.~\ref{pca1}~(d)).

In Fig.~\ref{pca1} and in Fig.~\ref{pca2} we have displayed our various results. Let us start with Fig.~\ref{pca1}~(a) where the results after a dimension reduction to two dimensions are shown for the full simplex. Here the letters mark the nine Bell states (``B''), the $12$ states of the kernel vertices (``K'') and the remaining $72$ vertices of the polytope (``P'') and are given for a better orientation of the kind of ``projection''. One can clearly observe that the ``projection'' is chosen by the specific ML algorithm to present some Bell states in the corner of a triangle. Furthermore, one clearly observes the spacing due to the step choice of $\Delta=\frac{1}{18}$ since there are $19$ ``lines''. The polytope vertices are projected onto two ``lines''.

In Fig.~\ref{pca1}~(b) the result is displayed for the full data of the polytope $\mathcal{P}$ with the color marking (FREE=green, BOUND=red, SEP=black, UNKNOWN=grey). One observes that the FREE vectors occupy the full space whereas the BOUND, SEP, UNKNOWN vectors populate in this ordering less and less space as we would expect from the known structure within the simplex $\mathcal{W}$. In more details, the mean length of FREE is $1.47\pm0.55$, of BOUND is $1.22\pm0.42$ and of SEP including UNKNOWN is $0.99\pm0.37$, hence there is significant difference visible.

As a next step we solely considered the vectors labelled by FREE, Fig.~\ref{pca1}~(c). The procedure finds a square where the vertices of $\mathcal{P}$ populate with different occurrence the corners. Again our grading is clearly visible.

The most exciting result is obtained, also very independent of the ordering of the vectors, for the set of BOUND entangled vectors, Fig.~\ref{pca1}~(d). All $156.600$  states ``project' to only $133$ points forming a triangle without vertices. For better orientation we put there also the results for the $84$ vertices of $\mathcal{P}$, which are of course not BOUND entangled states, which themselves show a nice geometry: forming a triangle and the origin for the $12$ kernel vertices and a hexagon with the origin for the remaining $72$ vertices of $\mathcal{P}$. Moreover, we marked the points by white and black colors if they were detected also by the MUB criterion and the quasispin criterion, respectively. One observes, three rings for each of the three symmetric parts of the triangle and that the MUB criterion, in contrast to quasipin criterion, detects the extremal, i.e. outermost states. The numerical witnesses cover the full space.

Firstly, this is of course an unexpected substructure that has been revealed by the ML algorithms, and secondly, this allows us to construct novel extremal states. Let us take two of the outmost states, e.g. $\rho_{\textrm{extremal 1}}=\frac{5}{18} (P_{0,2}+P_{1,2}+P_{2,2})+\frac{2}{18} P_{0,0}+\frac{1}{18}P_{1,1}$ and $\rho_{\textrm{extrema 2}}=\frac{5}{18} (P_{0,2}+P_{1,2}+P_{2,2})+\frac{1}{18} P_{0,0}+\frac{2}{18}P_{1,1}$. We expect that we can move the state until the line state $\rho_{\textrm{line}}=\frac{1}{3}(P_{0,2}+P_{1,2}+P_{2,2})$ in the corner (denoted by ``K''). Indeed, the state  $\rho_{\textrm{extremal 1}}=(\frac{1}{3}-x) (P_{0,2}+P_{1,2}+P_{2,2})+2 x  P_{0,0}+x P_{1,1}$ or $\rho_{\textrm{extremal 2}}=(\frac{1}{3}-x) (P_{0,2}+P_{1,2}+P_{2,2})+2 x  P_{1,1}+x P_{0,0}$ approaches the vertex of the triangle as depicted in Fig.~\ref{pcafamilyB2}~(a) for $x=\frac{1}{18},\frac{1}{36},\dots$, for which the value of the positive eigenvalue approaches zero and the MUB-witness is violated by exactly $\frac{1}{x}$. Moreover, we find that the states approaching the maximal violation of the MUB-witness, while still being PPT, are states towards the polytope vertices that are entangled (``green dot'' in Fig.~\ref{pcafamilyB2}~(a)). Those states correspond to the point with the greatest distance from the MUB witness of family $\mathcal{B}_2$ (also displayed in Fig.~\ref{pcafamilyB2}~(b)as a ``green dot'').

A second question concerns whether we have found all bound entangled states within our grading. For that we have applied a dimension reduction, Fig.~\ref{pca2}, to the set of kernel and UNKNOWN states (a), kernel (b), UNKNOWN (c) vectors, independently, and in (d) to the UNKNOWN and BOUND vectors. Though per construction the UNKNOWN states are outside of the $\mathcal{K}$ in the projection the appear inside of $\mathcal{K}$ and seem to be quite homogeneous spread. Most important, we do not obtain the strong structure, Fig.\ref{pca1}~(d), if the UNKNOWN states are added to those labelled BOUND, thus this is a further hint that we found all bound entangled states within our grading.

\subsection{Dimension Reduction followed by classification}

\begin{figure}
\begin{center}
(a)\includegraphics[width=5cm,keepaspectratio=true]{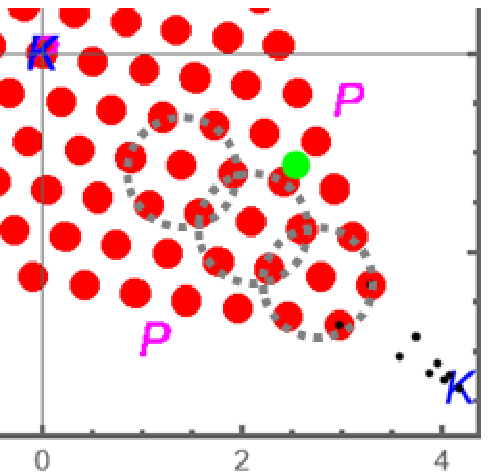}
(b)\includegraphics[width=5cm,keepaspectratio=true]{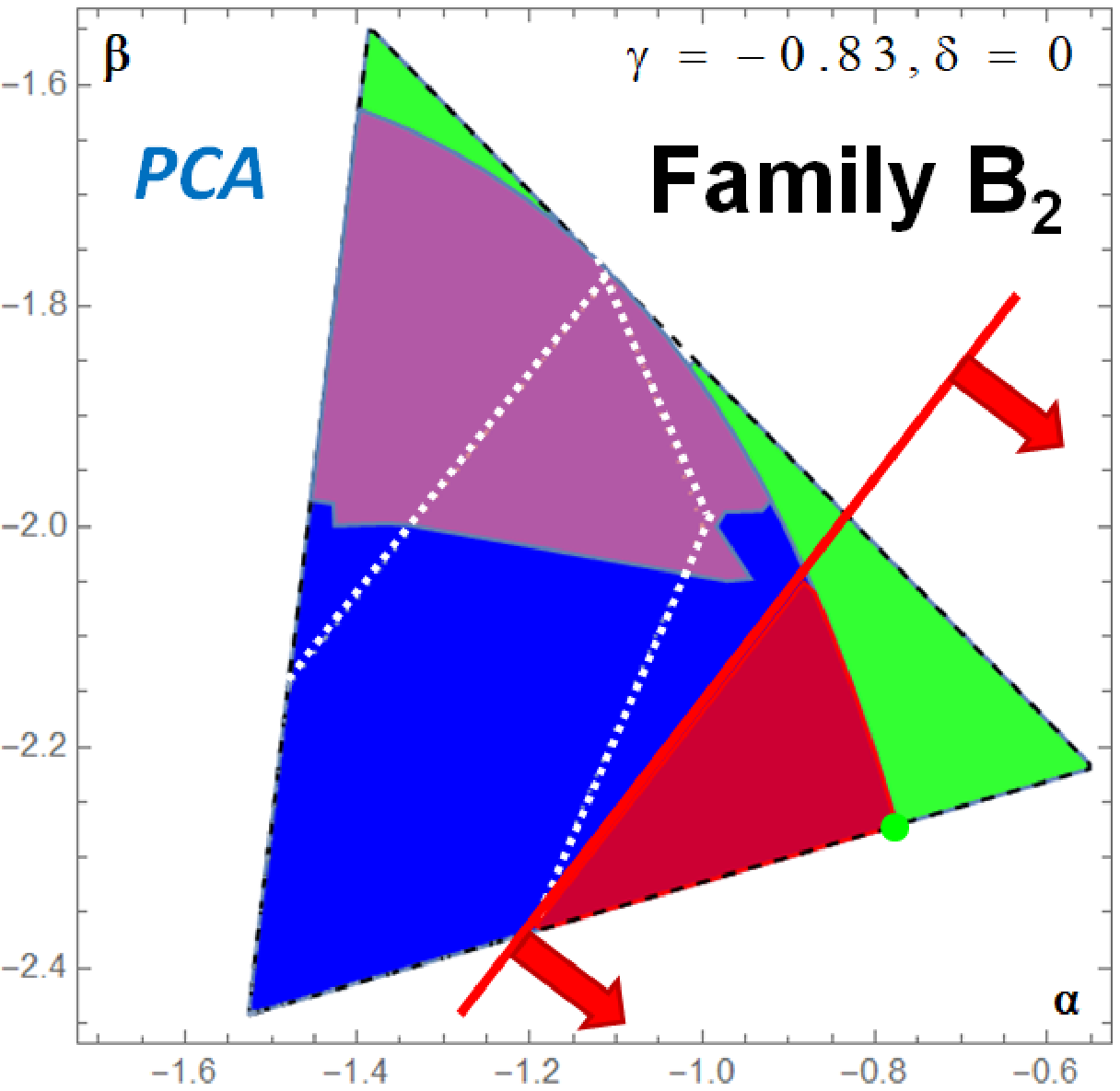}
\caption{In the graphic (a) the result of the dimensional reduction of the BOUND vectors are shown and how via the construction more bound entangled states can be deduced that converge to a vertex of $\mathcal{P}$, which is separable (denoted by ``K''). The rings show the states detected by the MUB witness. And the green dot corresponds to the maximal violation of the MUB-witness for the set of bound entangled states, which is found in the direction towards a vertex of $\mathcal{P}$, which is entangled (denoted by ``P'').  The graphic at the left hand side corresponds to a slice (family $\mathcal{B}_2$ via the simplex. The green point corresponds to the state state as depicted in (a). The purple area shows which states are detected to be bound entangled when the training data where the three dimensional vectors after the dimension reductions based on PCA.}\label{pcafamilyB2}
\end{center}
\end{figure}

A frequent method in ML is to reduce data by their principal components and then to apply classification methods. We have plugged in three principal component vectors into ML algorithms. For example, employing next-nearest algorithm we find an accuracy of $83\%$ and with the score probabilities $FREE\rightarrow 91\%, BOUND\rightarrow 55\%$ and $SEP\rightarrow 72\%$, which is lower than without reduction (Fig.~\ref{pcafamilyB2}~(b)). The result also suggests that there should be more bound entangled states than those detected by the analytical methods, however, also within the kernel region (``white dotted area'') which we know is wrong.

\section{Summary and Conclusion}

We discretized the Hibert space spanned by a complete set of maximally entangled bipartite states, the so called Bell states. We focussed on physical states with three degrees of freedom, i.e. bipartite qutrits. Those states are maximally mixed, meaning that tracing over one of the two subsystem results in a maximally mixed state, i.e. all results are equally probable independently of the measurement choices of Alice and Bob. Those magically symmetric states exhibit various symmetries, namely with respect to positivity it forms a simplex in $\mathcal{R}^{d^2-1}$. With respect to separability/entanglement one can construct a kernel polytope $\mathcal{K}$ for which every state inside has to be separable and an enclosure polytope $\mathcal{P}$ for which  every state outside is FREE entangled. Then we applied analytical and numerical entanglement witnesses, which let us finally label the vectors as FREE (free entangled), BOUND (bound entangled), SEP (separable) or if all methods failed as UNKNOWN, summarized in Fig.~\ref{simpelexsummary}.

Taking those results as the ground truth we investigated supervised and unsupervised ML algorithms with a threefold aim: (i) to recover new unnoticed symmetries (ii) to predict for a given state its probability of a certain label and (iii) to find out whether the UNKNOWN state are more likely separable or bound entangled.

Via clustering algorithms we find that assuming that the UNKNOWN states are SEP  is more consistent than the assumption BOUND (Fig.~\ref{confusionplots},) underpinning the fact that the set of separable states has to form a convex set, i.e. any convex combination of two separable states has to result in only separable states. Also the investigation of dimension reduction algorithms supported that we found all BOUND entangled states in the set (Fig.~\ref{pca1} and Fig.~\ref{pca2}).

The dimension reduction algorithms revealed a strong correlation in the set of BOUND vectors. All $156.600$ states are ``projected'' to $133$ points forming a triangle without vertices. This strong symmetry allowed us to construct novel bound entangled states in a systematic way, which are approaching the vertices belonging to the kernel. It will be interesting to see if this construction appears also in higher dimensions and if it depends on the explicit dimension.

Furthermore, we have chosen exemplary state families which turned out to be fruitful in developing the analytical methods due to particular properties. We used those data vectors, which are not necessarily those of our chosen exemplary families, to train various ML learning algorithms, which had quite high accuracies, however, they did not accurately produce the known symmetries nor always give reliable results in those chosen examples. We conclude from that more data vectors are needed for ML algorithms to classify better the simplex. Last but not least, we have combined the dimension reduction algorithms with the classifiers with a similar result.

It is often stated that only free entangled states are meaningful resources for quantum information tasks, which is certainly the case, when its performance relies on a good purity of the state under investigation. Therefore, a distillation protocol is often needed to compensate for experimental errors such as in long distant communications. Since only free entangled states are distillable, it is important to investigate decoherence channels that avoid a transformation of free entangled states into bound entangled states or separable states, such states have been named sudden death of distillability free~\cite{distilationsuddendeath}. With our achieved knowledge of the substructures in the magic simplex we can immediately deduce which ``trajectories'' lead to bound entangled regions and thereby single out the unital decoherence channels with and without sudden death.

In summary, we estimated the volumes of different entanglement types covered by a well defined family of states. Bound entangled states occur more often than separable states, even if the data vectors with the label UNKNOWN are all separable states. This simplex construction works in any dimension, thus the next steps are to discretize those in the same manner as presented in this contribution and in comparison with the results of this paper, one may disentangle effects from the dimension, the implied symmetries and  by that the ML learning algorithms may come up with a general novel method in detecting bound entanglement without explicit constructions of entanglement witnesses.

\textbf{Acknowledgement:}
B.C. Hiesmayr acknowledges gratefully the Austrian Science Fund (FWF-P26783). The author thanks for the fruitful discussions with Wojtek Krzemień, Dietmar Millinger, Franz Ramskogler and Wolfgang Waltenberger on ML algorithms and many thanks to Dan McNuty for carefully reading the manuscript. Furthermore, the Vienna Scientific Cluster (VSC) is acknowledged and the great support of their team.

\appendix

\section{PPT criterion}\label{pptcriterion}

A state $\rho$ is called PPT (positive partial transpose) iff the eigenvalues of the partial transpose with respect to one subsystem, $A$ or $B$, of a bipartition is positive:~\cite{PPT1,PPT2}
\beq
\label{PPT}
Tr_A\rho\geq 0,\;Tr_B\rho\geq 0
\eeq
If one or more eigenvalues are negative the state is (free) entangled. Only in low dimensions, i.e. $2\otimes 2$ and $2\otimes 3$, the PPT criterion is necessary and sufficient in detecting entanglement. Bound entangled states are PPT states and since the distillation is connected with the property of PPT~\cite{firstbound}, those states cannot be distilled into pure maximally (free) entangled states with local operations and classical communications.

\section{Quasipure criterion}\label{quasispincriterion}

One idea to approach the detection of entanglement of mixed states is to look for the closed pure state in the convex decomposition, a ``\textit{quasi-pure}'' approximation, and apply the generalized concurrence, which is certainly a lower bound on the total density matrix~\cite{quasipure}. A non-zero value detects entanglement. In detail, given a matrix $\rho=\sum_i p_i|\psi_i\rangle\langle\psi_i|$ the concurrence
\beq
C(\rho)=\inf_{\{p_i,|\psi_i\rangle\}} \sum_i p_i\; \langle \psi_i \psi_i|\mathcal{A}|\psi_i\psi_i\rangle^\frac{1}{2}
\eeq
with $\mathcal{A}= 4 \sum_{i<j,k<l}(|ikjl\rangle-|jkil\rangle-|iljk\rangle+|jlik\rangle).(\langle ikjl|-\langle jkil|-\langle iljk|+\langle jlik|)$ is bounded from
below
\beq
\label{quasispin}
C(\rho)&\geq& C_{qp}(\rho)=\max\left\{0,2 \max{\{\mathcal{S}_i\}_{i=0}^{d^2}}-\sum_{i} \mathcal{S}_i\right\}\;,
\eeq
where $\mathcal{S}_i$ are the singular values of $\mathcal{T}_{ij}=\sqrt{\mu_i\mu_j}\;\langle\psi_i\psi_j|\mathcal{A}|\psi_0\psi_0\rangle$ and $|\psi_0\rangle$ is the dominant eigenvector. For our magic simplex states, the dominant vectors are one of the Bell states, however, note that this criterion also sometimes fails, in particular if there is not a dominant eigenvector, e.g. for an equal mixture of two Bell states. The explicit non-linear witnesses for the family $A$ are published in Ref.~\cite{quasipure}.

\section{MUB-witnesses}\label{MUBcriterion}

Two orthonomal bases $\mathcal{B}_k=\{|i_k\rangle\}_{i=0}^{d-1},\mathcal{B}_l=\{|i_l\rangle\}_{i=0}^{d-1}$ are called mutually unbiased iff
\beq
|\langle i_k|j_l\rangle|^2&=&\delta_{k,l} \delta_{jl}+(1-\delta_{k,l})\cdot\frac{1}{d}\;.
\eeq
In the work~\cite{MUB1} it was shown that the observable
\beq
\mathcal{M}_m=\sum_{k=1}^{m}\sum_{i=0}^{d-1}|i_k i_k^*\rangle\langle i_k i_k^*|
\eeq
is an entanglement witness, more precisely, it has an upper bound $1+\frac{m-1}{d}$ for the set of all separable states, independent of the dimension. There is also a lower bound on this quantity by the set of separable states, however, it is generally depending on the dimension and on the choice of the set MUBs~\cite{DJDB1}. Note, that in particular to detect bound entangled states a reordering of the basis of one subsystem is needed else the witness $\mathcal{M}_m$ is decomposible. In this case the lower and upper bounds can change in general. There exists also unextendible MUBs~\cite{Danunext} which have been shown to more effective to detect entanglement but so far not bound entanglement. The strongest witness for $d=3$ is given for $m=d+1$, i.e. exploiting the complete set of MUBs which is not known to exist in any dimension. The slice $B_3$ is one where also $\mathcal{M}_d$ is successfully detecting bound entangled states~\cite{DJDB2}. Generally, the construction of entanglement witnesses from MUBs cover several well-known entanglement witnesses~\cite{Darek3} such as the Choi map.



%

\end{document}